\documentclass[reprint, superscriptaddress, amsmath, amssymb, prl]{revtex4-1}

\usepackage{graphicx}           
\usepackage{dcolumn}            
\usepackage{bm}                 
\usepackage{hyperref}   		
\usepackage{upgreek}
\usepackage{systeme}
\usepackage{url}
\usepackage{xcolor}
\usepackage{soul}

\begin{document}
	\title{Observation of the Dissipative Parametric Gain in a GaAs/AlGaAs 
		Superlattice}
	
	
	\author{Vladislovas~\v{C}i\v{z}as}
	\author{Liudvikas~Suba\v{c}ius}
	\author{Natalia V.~Alexeeva}
	\author{Dalius~Seliuta}
	\affiliation{Department of Optoelectronics, Center for Physical 
		Sciences and Technology, Saulėtekio Ave.~3, LT-10257, Vilnius, 
		Lithuania}
	\author{Timo Hyart}
	\affiliation{International Research Centre MagTop, Institute of 
		Physics, Polish Academy of Sciences, Al. Lotnik\'ow 32/46, 02-668 
		Warsaw, Poland}
	\affiliation{Department of Applied Physics, Aalto University, 00076 
		Aalto, Espoo, Finland}
	\author{Klaus K\"{o}hler}
	\affiliation{Fraunhofer-Institut f\"{u}r Angewandte Festk\"{o}rperphysik, Tullastrasse~72, Freiburg D-79108, Germany}
	\author{Kirill~N.~Alekseev}
	\email{kirill.alekseev@ftmc.lt (theory)}
	\affiliation{Department of Optoelectronics, Center for Physical 
		Sciences and Technology, Saulėtekio Ave.~3, LT-10257, Vilnius, 
		Lithuania}
	\affiliation{Department of Physics, Loughborough University, 
		Loughborough LE11 3TU, United Kingdom}
	\author{Gintaras~Valu\v{s}is}
	\email{gintaras.valusis@ftmc.lt (experiment)}
	\affiliation{Department of Optoelectronics, Center for Physical 
		Sciences and Technology, Saulėtekio Ave.~3, LT-10257, Vilnius, 
		Lithuania}
	\affiliation{Institute of Photonics and Nanotechnology, Department of 
		Physics, Vilnius University, Saulėtekio Ave. 3,
		LT-10257 Vilnius, Lithuania}
	
	
	\begin{abstract}
		Parametric generation of oscillations and waves is a paradigm, which is 
		known to be realized in various physical systems. Unique properties of 
		quantum superlattices allow to investigate high-frequency phenomena 
		induced by the Bragg reflections and negative differential velocity of 
		the miniband electrons. Effects of parametric gain in the superlattices 
		at different strengths of dissipation have been earlier discussed in a 
		number of theoretical works, but their experimental demonstrations are 
		so far absent. Here, we report on the first observation of the 
		dissipative parametric generation in a subcritically doped GaAs/AlGaAs 
		superlattice subjected to a dc bias and a microwave pump. We argue that 
		the dissipative parametric mechanism originates from a periodic 
		variation of the negative differential velocity. It enforces excitation 
		of slow electrostatic waves in the superlattice which provide a 
		significant enhancement of the gain coefficient. This work paves the 
		way for a development of a miniature solid-state parametric generator 
		of GHz\nobreakdash--THz frequencies operating at room temperature.
		
	\end{abstract}
	
	\maketitle
	
	\textit{Introduction.}--Parametric generation is a paradigm, known to be 
	realized in various physical systems ranging from electronic circuits up to 
	nonlinear optics. A large pump tone of the frequency $\omega_0$ causes a 
	periodic variation of a reactive element, which through mechanism of 
	parametric resonance results in the degenerate $2\omega_1=\omega_0$ or 
	nondegenerate $\omega_1+\omega_2=\omega_0$ processes of regenerative 
	amplification, and thus the both modes $1\&2$ can self-oscillate 
	\cite{Strutt1887, Rabinovich1989}. Positive gain can also be reached in the 
	phase-sensitive process of frequency up-conversion 
	$\omega_2+\omega_0=\omega_1$, but such amplification is not regenerative 
	and instead is governed by the familiar Manley-Rowe relations for powers 
	associated with each of the modes \cite{Manley1956}.
	
	However, there also exists less known dissipative parametric mechanism 
	associated with a periodic variation of a nonlinear resistance, and it is 
	often responsible for the generation of subharmonics in electric circuits 
	modelled by driven nonlinear oscillators \cite{Mandelstam1932}. This 
	mechanism requires that the system visits the state of negative 
	differential resistance during part of the ac pump period 
	\cite{Mandelstam1932, Migulin1983}. In the case of several modes, electric 
	powers associated with every frequency are connected by the Pantell 
	relations, which explicitly involve the differential conductance of 
	nonlinear resistive circuit \cite{Pantell1958}.
	
	Quantum semiconductor superlattices (SLs) \cite{Esaki1970} can be found as 
	a unique platform to meet the aforesaid condition. In these artificial 
	crystals, miniband electrons can perform electrically driven high-frequency 
	oscillations caused by the Bragg reflections \cite{Bass1986}. The major 
	focus is on the dc field induced Bloch oscillations, detectable both in 
	time \cite{Feldmann1992} and in space \cite{Lyssenko1997}, and on the 
	related dissipative phenomenon of Bloch gain \cite{Savvidis2004}. The 
	electron drift velocity ($v$) depends on the electric field following a 
	nonlinear curve that above a certain critical field ($E_\text{cr}$) 
	demonstrates the negative differential velocity (NDV) \cite{Esaki1970, 
	Sibille1990}. This active Esaki-Tsu nonlinearity is able to provide an 
	efficient multiplication of the microwave input frequency in SL-based 
	devices \cite{Ignatov1976}. It was a significant progress to realize such 
	THz frequency multipliers and mixers experimentally and obtain reasonable 
	power output suitable for various applications \cite{Renk2005, 
	Paveliev2012, Hayton2013}.
	
	\begin{figure*}
		\begin{center}
			\resizebox{\textwidth}{!}{
				\includegraphics{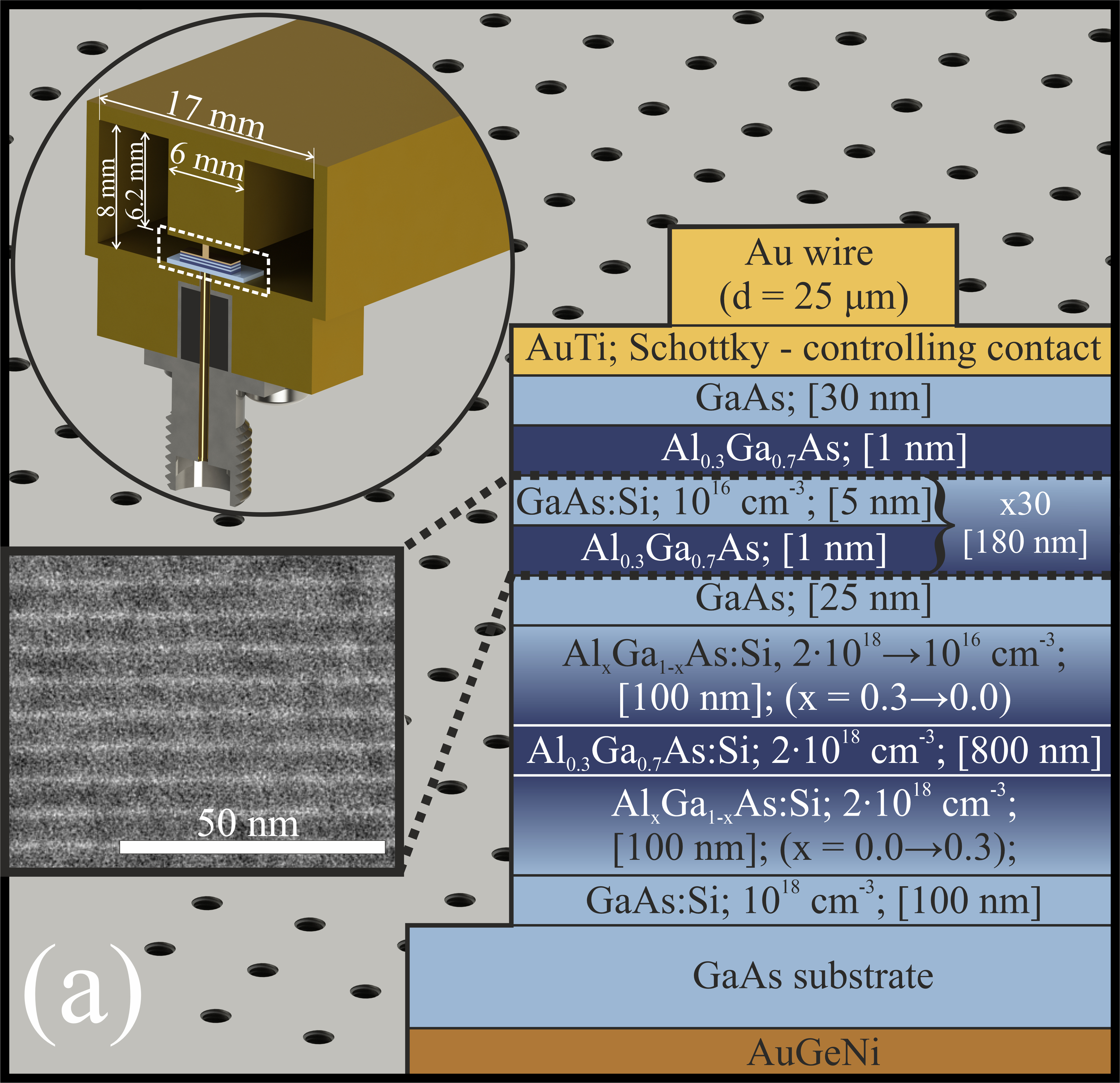}
				\includegraphics{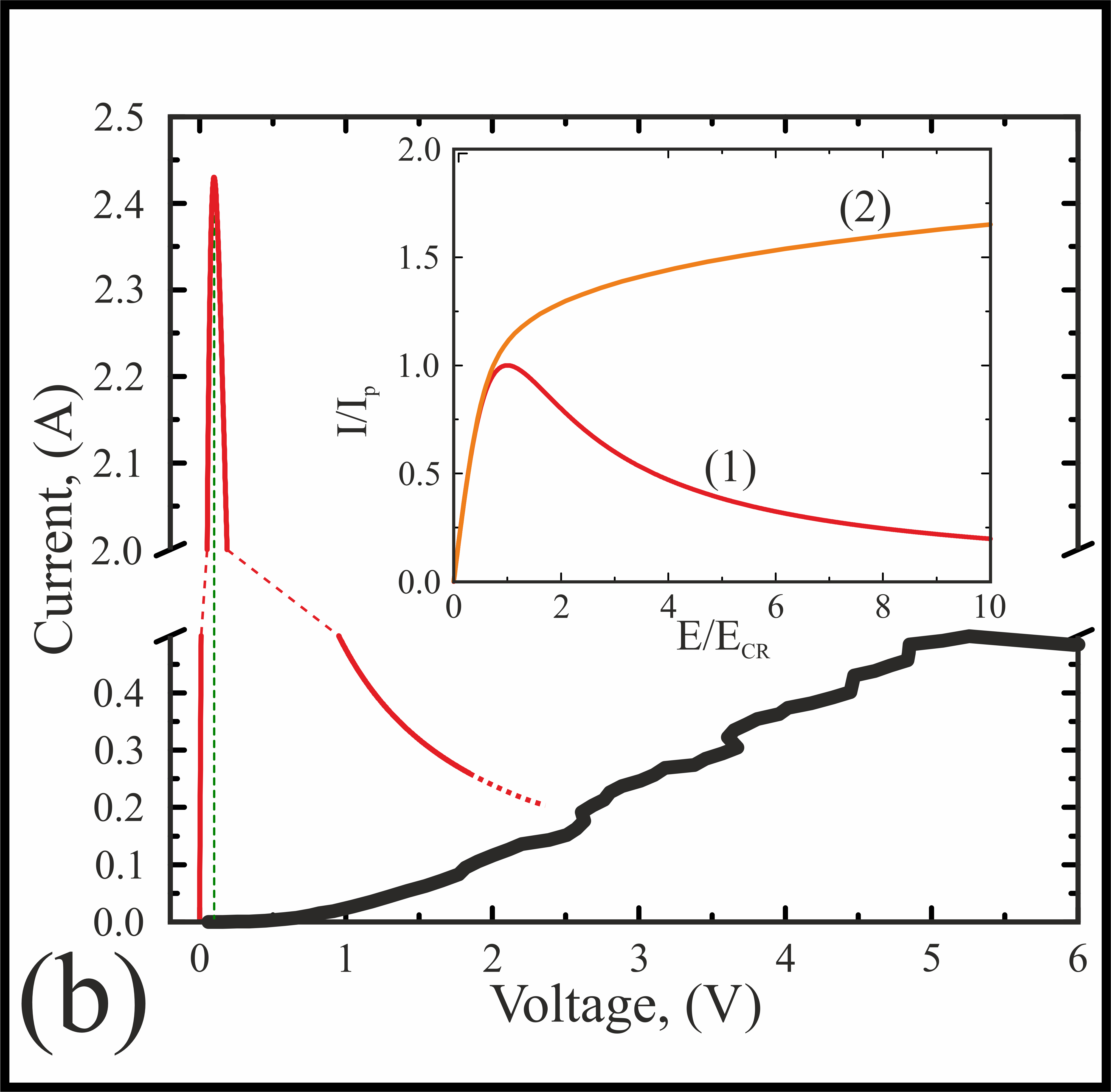}
				\includegraphics{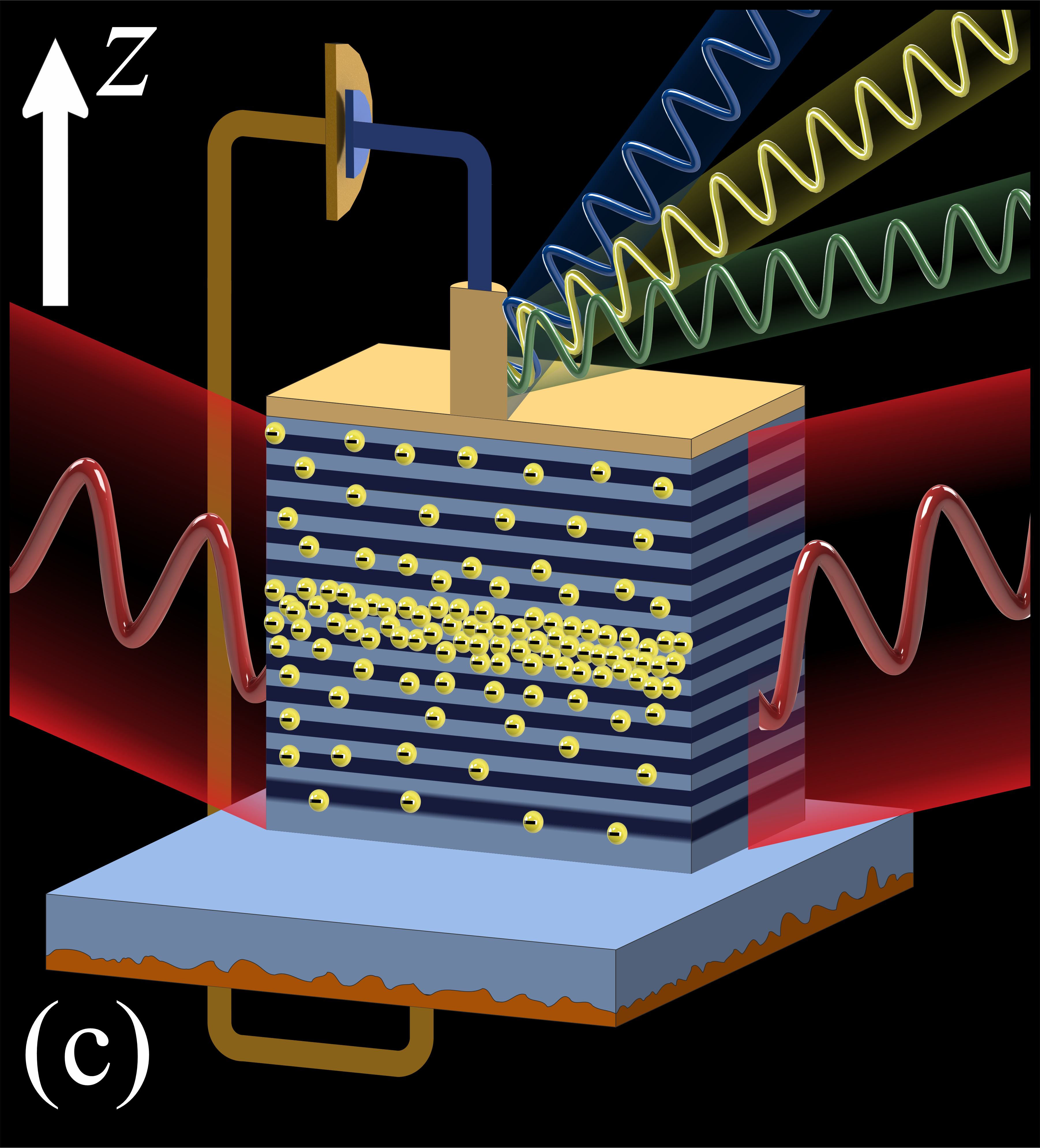}
			}
		\end{center}
		\caption{\label{fig:first}
			\textbf{(a):} Architecture of the GaAs/AlGaAs device where SL 
			sandwiched between the AuTi Schottky contact from the top and the 
			heterojunction from the bottom. To deliver the incident microwave 
			to the structure, it was placed along the wide wall of a 
			single-ridged waveguide of $17\times8$ mm. Inset: TEM image of the 
			superlattice.
			\textbf{(b):} The measured $I$-$V$ characteristic of the SL device 
			(thick black) in comparison with the ideal Esaki-Tsu $I$-$V$ 
			characteristic (thin red). The vertical (dashed green) line marks 
			the critical voltage ($E_\text{cr}L\approx 0.1$ V), for which the 
			peak current $I_p\approx 2.4$ A is reached. Inset: The Esaki-Tsu 
			curve (1, red) versus the calculated $I$-$V$ characteristic of SL 
			with ohmic injecting contact (2, orange). Notice that the 
			experimental $I$-$V$ curve runs significantly below $I_p$, 		
			which is opposite to the behaviour of $I$-$V$ characteristics in 
			the case of ohmic contacts. This points to nonohmic nature of 
			contacts in our SL device. 
			\textbf{(c):} Sketch of the parametric generation in the SL device. 
			Strong electromagnetic wave is coupled with electron plasma of the 
			dc biased SL. The electric field component of this pump wave, 
			directed along the SL axis $z$, parametrically interacts with the 
			miniband electrons and excites a dozen of growing waves at 
			frequencies satisfying the parametric conservation relations for 
			photons [Eq.~(\ref{eq:parametric_laws})]. The generated 
			longitudinal electrostatic waves propagate along $z$ at the  
			electron drift velocity before being converted to radiation through 
			the wire antenna. Spatial modulation of the electron density 
			associated with only one of such slow waves is shown schematically.}
	\end{figure*}
	
	Theories of parametric amplification and generation of high frequencies in 
	quantum SLs exist for more than 40 years \cite{Pavlovich1977, Orlov1980, 
	Orlov1982, Hyart2007, Shorokhov2009, Hyart2009_2, Romanov2000, Hyart2006, 
	Alekseev2006, Romanov2006}, and include, along with the earlier 
	contributions \cite{Pavlovich1977, Orlov1980, Orlov1982}, the thorough 
	analysis of the degenerate \cite{Hyart2007, Shorokhov2009}, nondegenerate 
	\cite{Hyart2009_2} and multifrequency \cite{Romanov2000} parametric 
	processes. Overall, the amplification is caused by the electronic Bragg 
	reflections in the narrow energy band; it was predicted to be multiphoton 
	and to exist for a very wide range of the pump frequencies ($\omega_0$) 
	that covers significant parts of GHz and THz frequency domains 
	\cite{Hyart2007}. Nevertheless, microscopic physical mechanisms behind the 
	parametric gain were found to be distinct whether $\omega_0\tau > 1$ or 
	$\omega_0\tau\ll1$ ($\tau\simeq 200$ fs is a characteristic relaxation time 
	at room temperature \cite{Savvidis2004,Renk2005}). In the case of 
	high-frequency pump $\omega_0\tau > 1$, the parametric gain has its origin 
	in a periodic variation of the effective electron mass and specific kinetic 
	inductance of the miniband electrons \cite{Hyart2007, Hyart2006}. 
	Furthermore, in the limit of small dissipation ($\omega_0\tau \gg 1$), the 
	Manley-Rowe relations are satisfied \cite{Hyart2009_2}. 
	
	The behaviour is quite different in the limit of strong dissipation 
	($\omega_0 \tau \ll 1$), which corresponds to GHz pump frequencies 
	\cite{Hyart2009_2, Alekseev2006, Romanov2006}. Now not only down-conversion 
	but also up-conversion process can provide regenerative amplification 
	\cite{Orlov1982, Hyart2009_2}. This dissipative parametric gain in SLs 
	\cite{Hyart2009_2, Alekseev2006, Romanov2006}, while being in essence 
	similar to that in nonlinear resistors, still relies on a periodic 
	switching to the NDV state rather than on negative values of the 
	differential conductance as in \cite{Pantell1958}. Indeed, an account of 
	space charge effects in the basic transport model reveals that the static 
	differential conductance of the moderately doped quantum SL is always 
	positive \cite{Maksimenko2016}, in agreement with the earlier findings 
	\cite{Shockley1954, Kroemer1970}. Despite decades of fruitful theoretical 
	developments, there was no experimental evidence of the parametric gain 
	effects in SLs so far.
	
	In this Letter, we report on the first experimental observation of the 
	dissipative parametric generation in a SL device subjected to a dc bias and 
	a microwave pump. We detect both the degenerate and nondegenerate 
	multiphoton parametric processes, together with signatures of large-signal 
	gain effects. In this room temperature experiment, we use a special design 
	of the device composed of a wide miniband GaAs/AlGaAs SL with asymmetric 
	nonohmic contacts, allowing to make the electric field profile within the 
	active part of the structure more uniform. We find that net optical gain 
	for this parametric device can be almost thousand times greater than it 
	follows from the earlier estimate \cite{Hyart2007}, and argue that the 
	enhancement originates from slow propagation of the intrinsic electrostatic 
	waves inside the SL \cite{Ignatov1987}. 
	
	\textit{Superlattice design and characterization.}--Experiments were 
	performed using GaAs/AlGaAs SL structure given in Fig.~\ref{fig:first}(a). 
	It contains $30$ periods of $5~$nm GaAs:Si quantum wells $(10^{16} \, 
	\rm{cm}^{-3})$ separated by 1~nm 
	Al\textsubscript{0.3}Ga\textsubscript{0.7}As barriers to enable a wide 
	miniband of $104~$meV. The AuTi Schottky contact was fabricated on top of 
	the structure, while from the SL bottom, GaAs/AlGaAs heterojunction was 
	formed. The SL processed into a square mesa of $80 \, \times 80 \,\upmu$m 
	dimensions and of 1.3~$\upmu$m height using wet etching was then mounted 
	inside the standard single-ridged waveguide. The gold wire of 
	$\simeq25~\upmu$m was attached to the top contact of the SL to ensure 
	proper coupling to microwaves and to serve for biasing of the structure. 
	The bottom ohmic contact was connected to the coaxial line via filter. 
	Measurements were performed employing a waveguide-based setup 
	\cite{Subacius2015} relying on changes in the microwave transmission and 
	reflection induced by the electron transport in the SL \cite{Supplement}. 
	
	For this SL the product of the doping density ${N=10^{16} \, \rm{cm}^{-3}}$ 
	and the length ${L=180 \, \rm{nm}}$ is below the specific critical value 
	determining the onset of the travelling high-field domains, 
	${(NL)_{\rm{cr}}=7\epsilon E_{\rm{cr}}/e=2.7\times 10^{11} \, 
	\rm{cm}^{-2}}$ \cite{Kroemer1964, Ignatov1985, Supplement}. Therefore, our 
	subcritically doped SL operates in the electrically stable transport regime 
	\cite{Kroemer1965}, in a similar way to the well-known experiments 
	\cite{Sibille1989, Savvidis2004}. 
	
	The experimental $I-V$ characteristic of the SL device, measured employing 
	electrical pulses of 20~ns duration, is presented in 
	Fig.~\ref{fig:first}(b). It has positive slope, which is typical feature of 
	electronic systems with NDV operating in the stable transport regime 
	\cite{Kroemer1970, Dascalu1974,Maksimenko2016}. By taking this for granted, 
	we find how a comparison of the experimental and so-called neutral $I-V$ 
	characteristics can be used to extract information on the nature of 
	electric contacts of the SL. The neutral characteristic references to a 
	special situation of the electric neutrality, when densities of the mobile 
	and fixed ($N$) charges coincide, electric field is homogeneous and 
	contacts are absent \cite{Dascalu1974}. In our case, this is the ideal 
	Esaki-Tsu $I-V$ characteristic, the shape of which directly follows the 
	$v(E)$ dependence. Next, both the calculations \cite{Maksimenko2016} and 
	the experiments \cite{Sibille1989, Savvidis2004} expose that $I-V$ 
	characteristics of SLs with ohmic contacts typically saturate either above 
	the Esaki-Tsu peak current $I_p$ [Fig.~\ref{fig:first}(b) inset] or on the 
	level of $I_p$ [inset of Fig. 1 in \cite{Savvidis2004}]. On the contrary, 
	the measured $I-V$ characteristic runs significantly below $I_p$ 
	[Fig.~\ref{fig:first}(b)], indicating the nonohmic nature of contacts in 
	our device. We attribute this nonohmicity to the presence of Schottky and 
	shallow heterojunction barriers. Typically, the use of nonohmic contacts in 
	NDV devices makes the electric field profile more uniform, and thus 
	contributes to the better performance \cite{Yu1971, Dascalu1974}. In 
	addition, the built-in voltage of 0.65~V increases the voltage drop across 
	the SL by this amount. 
	
	\textit{Experimental results and discussion}. -- Experiment on the 
	parametric generation in the SL device is sketched in 
	Fig.~\ref{fig:first}(c). Strong electromagnetic wave passes through the SL 
	along its layers, modulates the electron differential velocity, and by 
	means of the dissipative parametric mechanism excites a spectrally rich 
	coherent emission from the device into the output waveguide. In the earlier 
	theoretical suggestions, the generated spectral components were assumed to 
	be transverse electromagnetic modes of the external cavity (cf. Fig.~1 in 
	\cite{Hyart2007}). Contrastingly, our device operates without such 
	resonator and mainly relies on intrinsic longitudinal modes inside the SL 
	that are growing from fluctuations of the electron plasma. We will return 
	to the discussion of the origin and significance of these intrinsic modes 
	after consideration of the generated frequencies in the light of the major 
	predictions of the theory.
	
	\begin{figure}[b]
		\includegraphics[width=8.6cm, keepaspectratio]{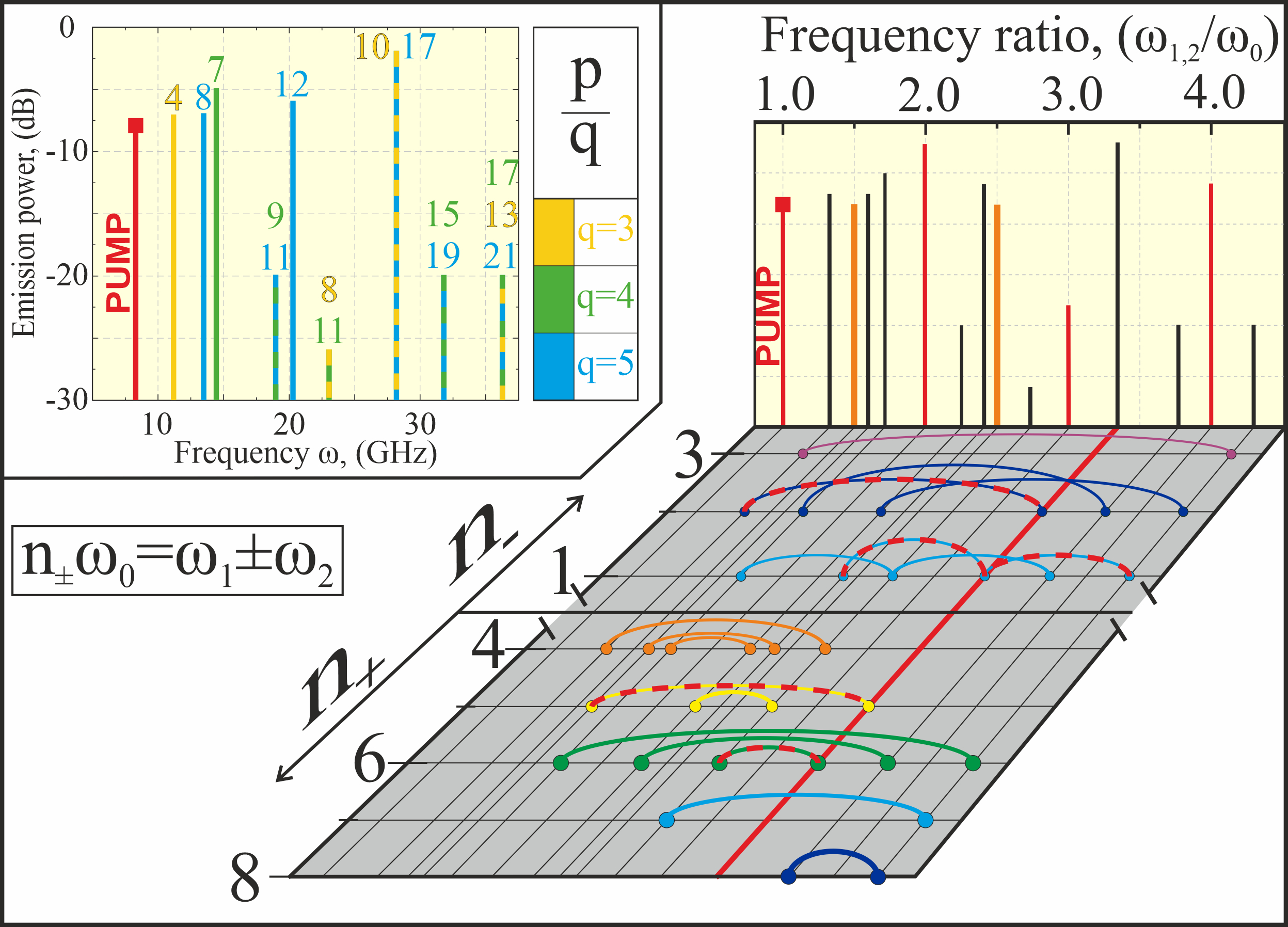}
		\caption{\label{fig:nonlinearity} The spectrum of the frequencies 
		generated in SL under the action $8.45$~GHz pump and bias of 
		$0.3$~V. The frequencies are scaled to the pump frequency (vertical 
		panel). The spectral lines corresponding to the pump and its 
		multiplication (red), the sub-harmonics (orange), and to the 
		nondegenerate parametric processes (black) are displayed. The 
		$1/2$-harmonic is not visible due to cut-off characteristics of the 
		output waveguide. The links in the horizontal panel indicate that the 
		observed emission lines follow the spectrum of the small-signal 
		parametric gain [Eq.~(\ref{eq:parametric_laws})]. Multiphoton and 
		multiple wave mixing phenomena (red dashed lines), related to the 
		generation at $28.20~\rm{GHz}$ (horizontal solid red line), are further 
		exemplified in Eqs.~(\ref{eq:cascasding_v2}). 
		Inset: The frequencies classified according to the pump fractions 
		[Eq.~(\ref{eq:frac_freq})]. The corresponding numerators $p$ are shown 
		nearby their spectral lines, while the denominators $q$ are defined 
		following the color chart.}
	\end{figure}
	
	The measured emission spectrum for the case of $8.45$~GHz pump is presented 
	in Fig.~\ref{fig:nonlinearity} (vertical panel). The strength of ac field 
	inside the SL was estimated to be $\simeq 8E_\text{cr}$ \cite{Supplement}. 
	Along with the pump and its harmonics up to the $4^{th}$ order, the 
	spectrum also contains additional 11 emission lines. The prerequisite for 
	an enforcement of the stimulated emission at these discrete frequencies is 
	a positive gain for infinitesimal signals. The theory states that the 
	small-signal parametric gain in SLs can arise only for the frequencies 
	$\omega_{1,2}$ that are connected to the pump frequency as 
	\begin{equation}
		\label{eq:parametric_laws}
		\omega_1\pm \omega_2=n_\pm\omega_0, \quad 2\omega_1=n_0\omega_0, 
	\end{equation}
	\begin{figure*}
		\begin{center}
			\resizebox{\textwidth}{!}{
				\includegraphics{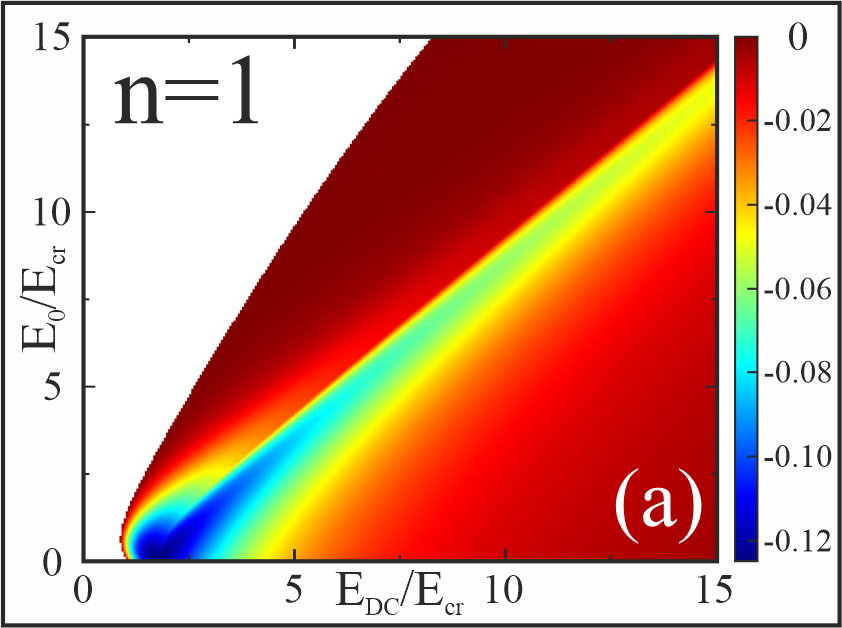}
				\includegraphics{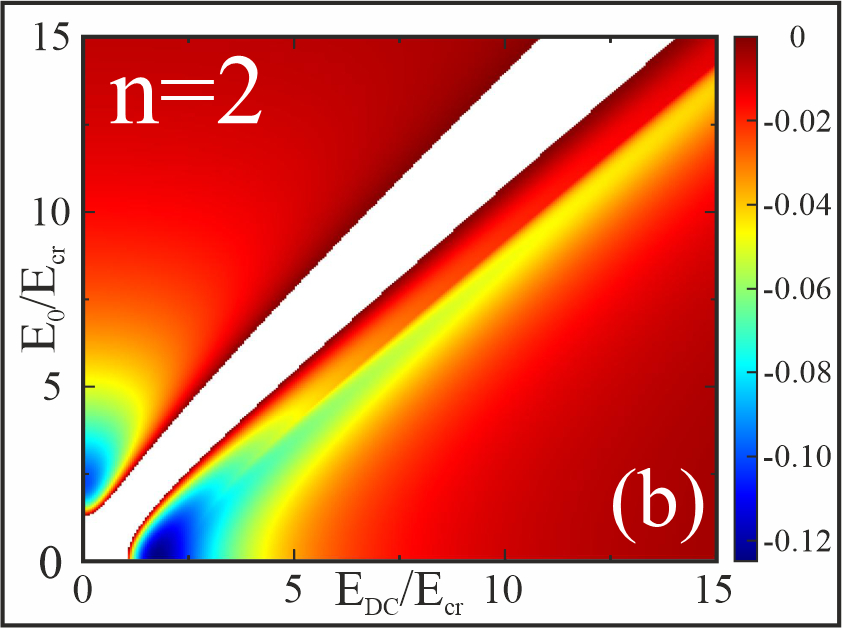}
				\includegraphics{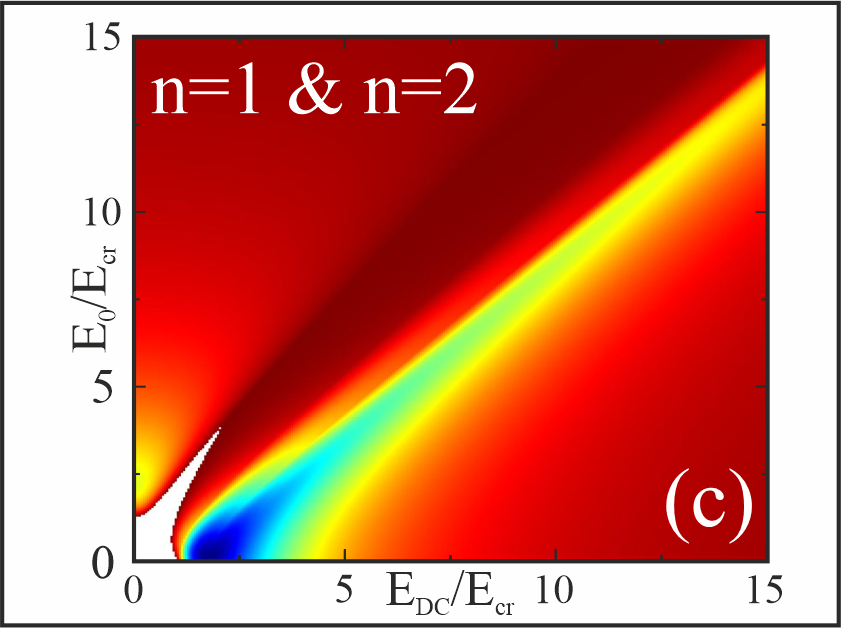}
			}
		\end{center}
		\caption{\label{fig:second}
			\textbf{(a-b):} The areas and magnitudes of gain $\mu_n<0$ (color) 
			in the plane of dc bias $E_\text{dc}$ and pump field amplitude 
			$E_0$ for $n=1$ (a) and $n=2$ (b). The calculated high-frequency 
			mobilities $\mu_n$ are presented in units of the superlattice Drude 
			mobility ($2v_p/E_\text{cr})$, and the dc and ac electric fields 
			are scaled to the critical field $E_\text{cr}$. Blank areas 
			everywhere correspond to absorption ($\mu_n>0$).
			\textbf{(c):} Overlapping of the areas corresponding to both 
			$\mu_1<0$ and $\mu_2<0$ (color) in the same plane. The figure 
			illustrates an overall extension of the gain area with the only 
			boundary at low fields, in the case when two parametric processes 
			are simultaneously realised in the SL. 
		}
	\end{figure*}
	where $n_+$, $n_-$ are positive integers and $n_0$ is odd 
	\cite{Hyart2007,Hyart2009_2, Romanov2006}. The existence of the parametric 
	relations with $n\neq 1$ is a definitely notable property of the quantum 
	superlattice nonlinearity. Whereas generation of the half-harmonics and 
	spontaneous down-conversions with various $n_{+}$ are universal signatures 
	of the parametric gain in SLs \cite{Hyart2007}, the appearance of 
	self-oscillations at both frequencies involved in the up-conversion 
	processes [Eq.~(\ref{eq:parametric_laws}) with $n_{-}\geq 1$] is a 
	remarkable property of the dissipative mechanism 
	\cite{Orlov1982,Hyart2009_2}. 
	
	The data and links displayed in Fig.~\ref{fig:nonlinearity} affirm that the 
	observed parametric emission lines (black and orange) satisfy the photon 
	energy conservation relations of Eqs.~(\ref{eq:parametric_laws}), and thus 
	can include many photons of the pump $(n>1)$. Surprisingly, we also found 
	that almost every spectral line participates in several multiphoton 
	processes simultaneously. For instance, the frequency $28.20$~GHz is 
	generated in the following processes (cf.~Fig.~\ref{fig:nonlinearity}):
	\begin{equation} 
		\label{eq:cascasding_v2}
		\begin{cases}
			\,\textbf{28.20} + 23.05 = 6\times 8.45\,\text{GHz} \,\text{(8-photons)}, \\
			\,\textbf{28.20} + 13.45 = 5\times 8.45\,\text{GHz} \,\text{(7-photons)}, \\
			\,\textbf{28.20} - 11.20 = 2\times 8.45\,\text{GHz} \,\text{(4-photons)}, \\
			\,\textbf{28.20} - 20.30 = 1\times 8.45\,\text{GHz} \,\text{(3-photons)}, \\
			\,36.25 - \textbf{28.20} = 1\times 8.45\,\text{GHz} \,\text{(3-photons)}. \\
		\end{cases}
	\end{equation}
	This unusual behaviour in SLs is in sharp contrast to the parametric 
	generation in conventional optical systems, where multiphoton effects 
	involving the pump ($n\hbar\omega_0$) are virtually absent, and even 
	multi-step frequency cascades \cite{Andrews1970} are well described by the 
	Manley-Rowe relations \cite{Chirkin2000, Saltiel2005}. As a consequence, 
	optical parametric generation of the frequencies involved in the 
	up-conversion processes with several pump photons ($n_-=2,3$) is unlikely 
	realizable \cite{Saltiel2005}, but such emission lines are readily 
	observable in the SL, see Fig.~\ref{fig:nonlinearity}.
	
	Furthermore, the measured emission spectrum can bear signatures of 
	generation effects that go beyond the linear response in signal strengths. 
	In the case of large signals ($\simeq E_\text{cr}$), it is predicted that 
	new channels in parametric generation induce the fractional frequencies 
	\begin{equation}
		\label{eq:frac_freq}
		\omega_1 = \left(p/q\right)\omega_0, \quad q>2,
	\end{equation}
	where $p$, $q$ are integers \cite{Shorokhov2009, Romanov2000}. In 
	particular, the subharmonics $p\omega_0/3$ and $p\omega_0/4$ can arise due 
	to effects that are quadratic and cubic in the signal strength, 
	respectively \cite{Shorokhov2009}. Analysis of the data 
	(Fig.~\ref{fig:nonlinearity}~inset) unveils this type of frequencies and 
	thus experimentally confirms involvement in the large signal regime. 
	Similar emission spectra, the spectral components of which do fit 
	Eqs.~(\ref{eq:parametric_laws})~and~(\ref{eq:frac_freq}), were observed for 
	other pump frequencies close to 10~GHz. Different multiphoton parametric 
	processes in SL can be also distinguished by comparing their input-output 
	power dependencies \cite{Supplement}. At the present state of the art, 
	however, this method is less informative than the used spectroscopic 
	approach.
	
	\textit{Theoretical justification}.--To get deeper insight into the role of 
	the observed multiple simultaneous parametric processes, we calculate and 
	compare the high-frequency electron mobilities $\mu_n(E_\text{dc}, E_0)$ 
	for two processes with $n=1$ and $n=2$ separately 
	[Eq.~(\ref{eq:parametric_laws})], and also for their combination [cf. Eqs. 
	(\ref{eq:cascasding_v2})]. The total pump electric field was assumed to be 
	the sum of dc bias $E_\text{dc}$ and strong ac field $E_0\cos(\omega_0 t)$, 
	where $\omega_0\tau\ll 1$ \cite{Supplement}. The calculated areas of gain 
	($\mu_n<0$) and absorption ($\mu_n>0$) in the plain $E_0$-$E_\text{dc}$ are 
	presented in Fig.~\ref{fig:second}. For the both single parametric 
	processes shown in the subplots (a)~and~(b), there exist wide blank areas 
	of no-gain, which effectively can cause appearance and disappearance of the 
	amplification at relatively large values of either $E_0$ or $E_\text{dc}$. 
	However, in the case the two parametric processes run simultaneously, the 
	gain area expands and the blank area survives only for rather weak applied 
	fields $\alt E_\text{cr}$ [Fig.~\ref{fig:second}(c)]. This determines a 
	well-defined threshold line for the positive gain when $E_0\simeq 
	E_\text{dc}\simeq E_\text{cr}$. Therefore, an account of the multiple 
	processes in SLs restores intuitive correspondence with the condition of 
	dissipative parametric amplification in nonlinear resistors.
	
	We turn to the origin of the intrinsic electrostatic modes 
	[Fig.~\ref{fig:first}(c)] and their contribution to the net optical gain. 
	Generally, every such electrostatic mode propagates at the drift velocity 
	of electrons \cite{Lampert1956}, and represents an undamped excitation of 
	the solid state plasma in condition of NDV \cite{Schoell1987}. Specifically 
	for the Esaki-Tsu active nonlinearity, it is known as the drift-relaxation 
	self-mode of SL \cite{Ignatov1987} caused by the Bragg reflections of the 
	miniband electrons \cite{Buttiker1977}. We apply this concept to the case 
	of the dissipative parametric gain.
	
	Consider an intrinsic longitudinal mode which frequency satisfies one of 
	the parametric relations of Eqs.~(\ref{eq:parametric_laws}) at some fixed 
	photon number $n$. If the corresponding $\mu_n<0$, in the linear stage 
	small fluctuations of the electric field grow exponentially while electron 
	flow propagates with the drift velocity $v$ through the sample of the 
	length $L$. The growth rate of the electrostatic wave is determined by the 
	product of the dielectric relaxation frequency $(eN\mu_n)/\epsilon$ and the 
	electron transit time \textit{L/v} \cite{Heinle1968}. Therefore, the gain 
	coefficient $\beta_n$, defined by means of the Beer law 
	$I_\text{out}=I_\text{in}\exp(-\beta z)$, can be estimated as 
	\begin{equation}
		\label{eq:opt-gain}
		\beta_n=\frac{2eN\mu_n}{\epsilon v}.
	\end{equation}
	Remarkable, the direct substitution $v/2\simeq v_p\rightarrow c'$ 
	transforms the right-hand side of Eq.~(\ref{eq:opt-gain}) into the 
	corresponding gain coefficient of the electromagnetic mode travelling at 
	speed of light in the semiconductor $c'=c/n_r$ ($n_r$ is the average 
	refractive index of SL) \cite{Wacker2002, Hyart2007}. Since 
	$v_p/c'\simeq~10^{-3}$ contribution of the slow electrostatic modes to the 
	net optical gain can prevail. By assuming $\mu_1/\mu_0\approx~-0.02$ for 
	the 3-photon parametric processes [Fig.~\ref{fig:second}(a)] we obtain from 
	Eq.~(\ref{eq:opt-gain}) large gain $\beta_1\gtrsim10^4$~cm$^{-1}$ for the 
	corresponding slow modes.
	
	Finally, the generated electrostatic modes are transformed to the coherent 
	electromagnetic radiation through a wire antenna bonded to the SL 
	[Fig.~\ref{fig:first}(c)]. It is worth noticing that similar 
	phase-preserving conversions of electromagnetic waves to plasmons and back 
	have been demonstrated in nanometric field effect transistors 
	\cite{Drexler2012, Matyushkin2020}.
	
	\textit{Conclusion.}-We observed and explained unusual parametric 
	generation in a quantum optoelectronic system with strong dissipation when 
	the Manley-Rowe relations are broken. We showed that intensive microwave 
	pumping of the GaAs/AlGaAs superlattice stimulates dissipative parametric 
	gain at room temperature, manifesting itself as a steady coherent emission 
	at various fractional harmonics of the pump frequency. Unusually, this 
	device starts to self-oscillate in up-conversion processes as easy as in 
	down-conversion ones. We also revealed the significance of the undamped 
	drift-relaxation modes for the amplification mechanism in the superlattice. 
	These slow plasma waves can provide large gain of $\simeq 10^4$ cm$^{-1}$ 
	and more, thus enabling multiphoton generation the in cavityless 
	configuration. Our experiments confirmed core predictions of the existing 
	theory of dissipative parametric amplification at GHz frequencies, and also 
	further stretched its limits by describing the multiple parametric 
	processes. At once, semiconductor quantum SLs hold the promise of room 
	temperature parametric amplification in the technologically important 
	sub-THz range and beyond \cite{Hyart2007}.
	
	\begin{acknowledgments}
		We are sincerely grateful to Martynas Skapas, Sandra Stanionyt\.{e} and 
		Remigijus Ju\v{s}k\.{e}nas for superlattice characterization; Linas 
		Minkevi\v{c}ius, Vladimir Maksimenko, and Miron~S.~Kagan for 
		illuminating discussions. Research activities of K.N.A. were partially 
		funded by Marius Jakulis Jason Foundation and T.H. was supported, in 
		part, by Foundation for Polish Science through the IRA Programme 
		co-financed by EU within SG~OP.
	\end{acknowledgments}

\end{document}


\title{Supplemental Material\\ 
		Observation of the Dissipative Parametric Gain in a GaAs/AlGaAs 
		Superlattice}
	\author{Vladislovas~\v{C}i\v{z}as}
	\author{Liudvikas~Suba\v{c}ius}
	\author{Natalia V.~Alexeeva}
	\author{Dalius~Seliuta}
	\affiliation{Department of Optoelectronics, Center for Physical Sciences 
		and Technology, Saulėtekio Ave.~3, LT-10257, Vilnius, Lithuania}
	\author{Timo Hyart}
	\affiliation{International Research Centre MagTop, Institute of 
		Physics, Polish Academy of Sciences, Al. Lotnik\'ow 32/46, 02-668 
		Warsaw, Poland}
	\affiliation{Department of Applied Physics, Aalto University, 00076 
		Aalto, Espoo, Finland}
	\author{Klaus K\"{o}hler}
	\affiliation{Fraunhofer-Institut f\"{u}r Angewandte Festk\"{o}rperphysik, 
		Tullastrasse~72, Freiburg D-79108, Germany}
	\author{Kirill~N.~Alekseev}
	\affiliation{Department of Optoelectronics, Center for Physical 
		Sciences and Technology, Saulėtekio Ave.~3, LT-10257, Vilnius, 
		Lithuania}
	\affiliation{Department of Physics, Loughborough University, 
		Loughborough LE11 3TU, United Kingdom}
	\author{Gintaras~Valu\v{s}is}
	\affiliation{Department of Optoelectronics,
		Center for Physical Sciences and Technology, Saulėtekio Ave.~3, 
		LT-10257, Vilnius, Lithuania}
	\affiliation{Institute of Photonics and Nanotechnology, 
		Department of Physics, Vilnius University, Saulėtekio Ave. 3,
		LT-10257 Vilnius, Lithuania}
	\maketitle
	\vspace{-1.5cm}
	\tableofcontents
	
	\section{Superlattice design and experimental setup}
	
	\begin{figure*}[!h]
		\centering
		\includegraphics[width=\textwidth]{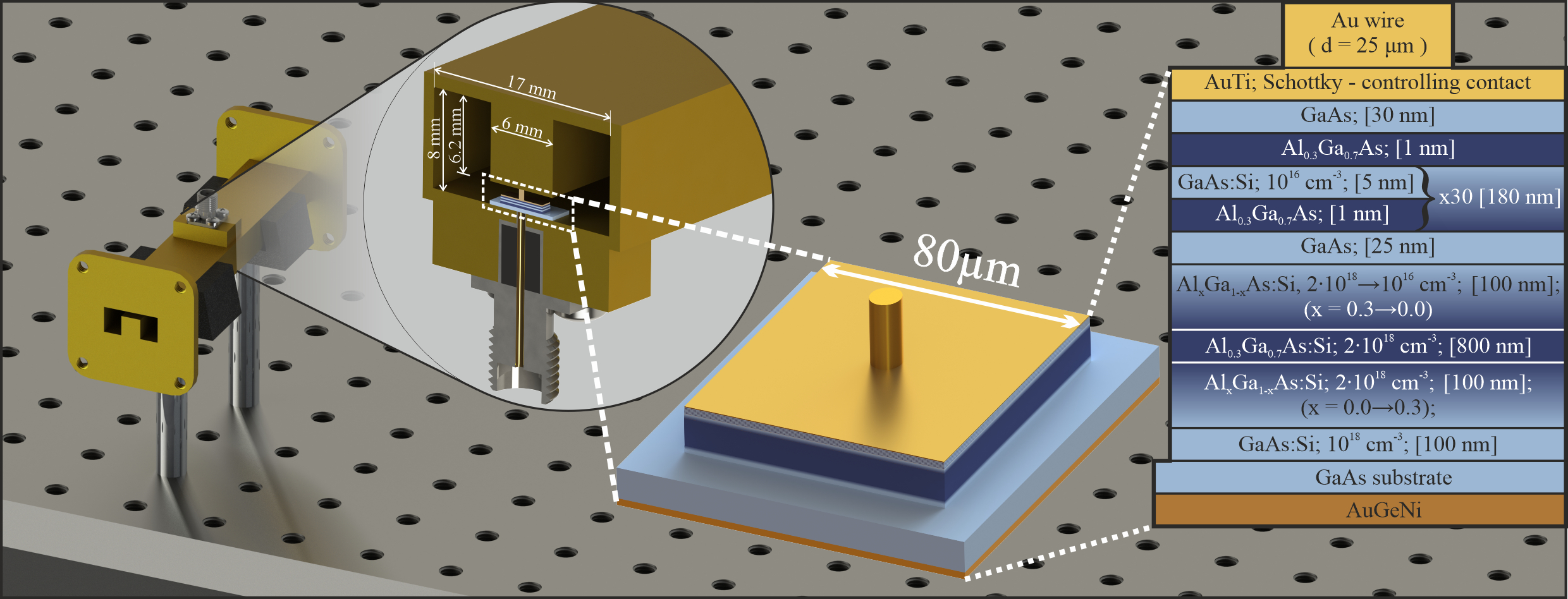}
		\caption{\label{fig:structure} Part of the measurement setup with 
		enlarged view of the sample holder for mounting the structure. The 
		sample was placed along the wide wall of a single-ridged waveguide with 
		dimensions of $\rm{17\times8\,mm}$ to guide the incident microwave 
		radiation. To provide the effective coupling of the radiation into the 
		structure, $\rm{25\,\upmu m}$ diameter golden wire was bonded to the 
		top contact serving also for application of the bias voltage over the 
		SL. Bottom Ohmic contact of the SL was connected to coaxial line via 
		filter. Total thickness of the grown structure was 
		$\sim1.3\,\upmu\rm{m}$; rectangular mesa of $80\times80\,\upmu\rm{m}$ 
		was processed.}
	\end{figure*}
	
	Specially designed GaAs/AlGaAs superlattice (SL) was grown using molecular 
	beam epitaxy technique on $n$-doped GaAs substrate $\rm{(\sim500~\upmu m; 
	10^{18}~cm^{-3})}$. The architecture of the SL structure is given in the 
	right panel of Fig.~\ref{fig:structure}. It contains 30 periods of 
	moderately doped quantum wells GaAs:Si $(\rm{5~nm; 10^{16}~cm^{-3}})$ 
	separated by $\rm{Al_{0.3}Ga_{0.7}As}$ (1~nm) barriers and enabling thus a 
	wide miniband (104~meV) in the SL. AuTi Schottky contact, following GaAs 
	buffer layer (30~nm), were fabricated on the top of the structure. From the 
	bottom side of the SL, GaAs (25~nm) layer with $\rm{Al_xGa_{1-x}As}$ 
	heterojunction was formed (100~nm; gradient-type alloy and concentration; 
	for more details see inset in Fig.~\ref{fig:structure}). The bottom contact 
	on the substrate was Ohmic, realized by annealing of AuGeNi alloy.
	
	\begin{figure*}[h]
		\begin{center}
			\resizebox{\textwidth}{!}{
				\includegraphics{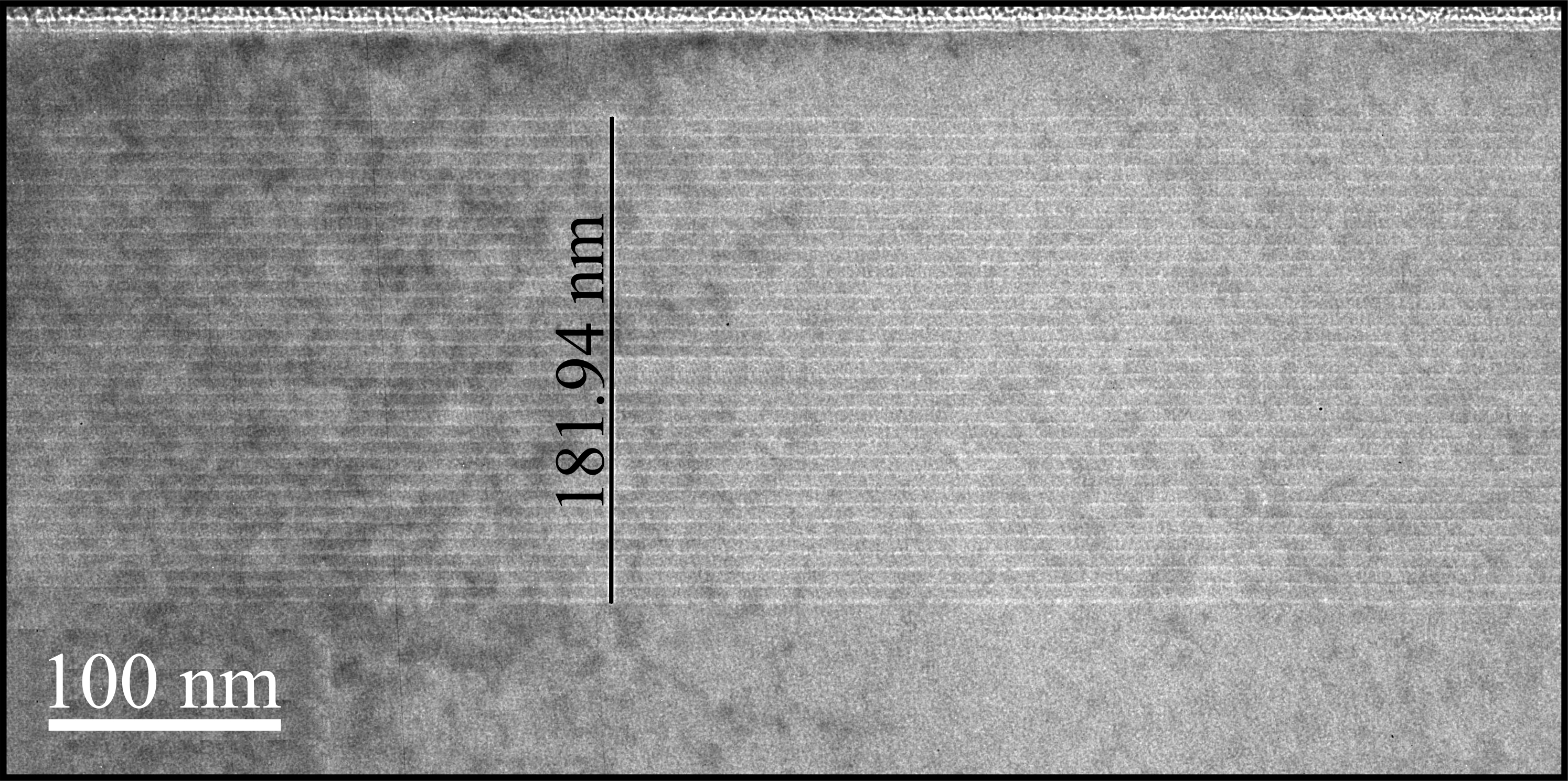}
				\includegraphics{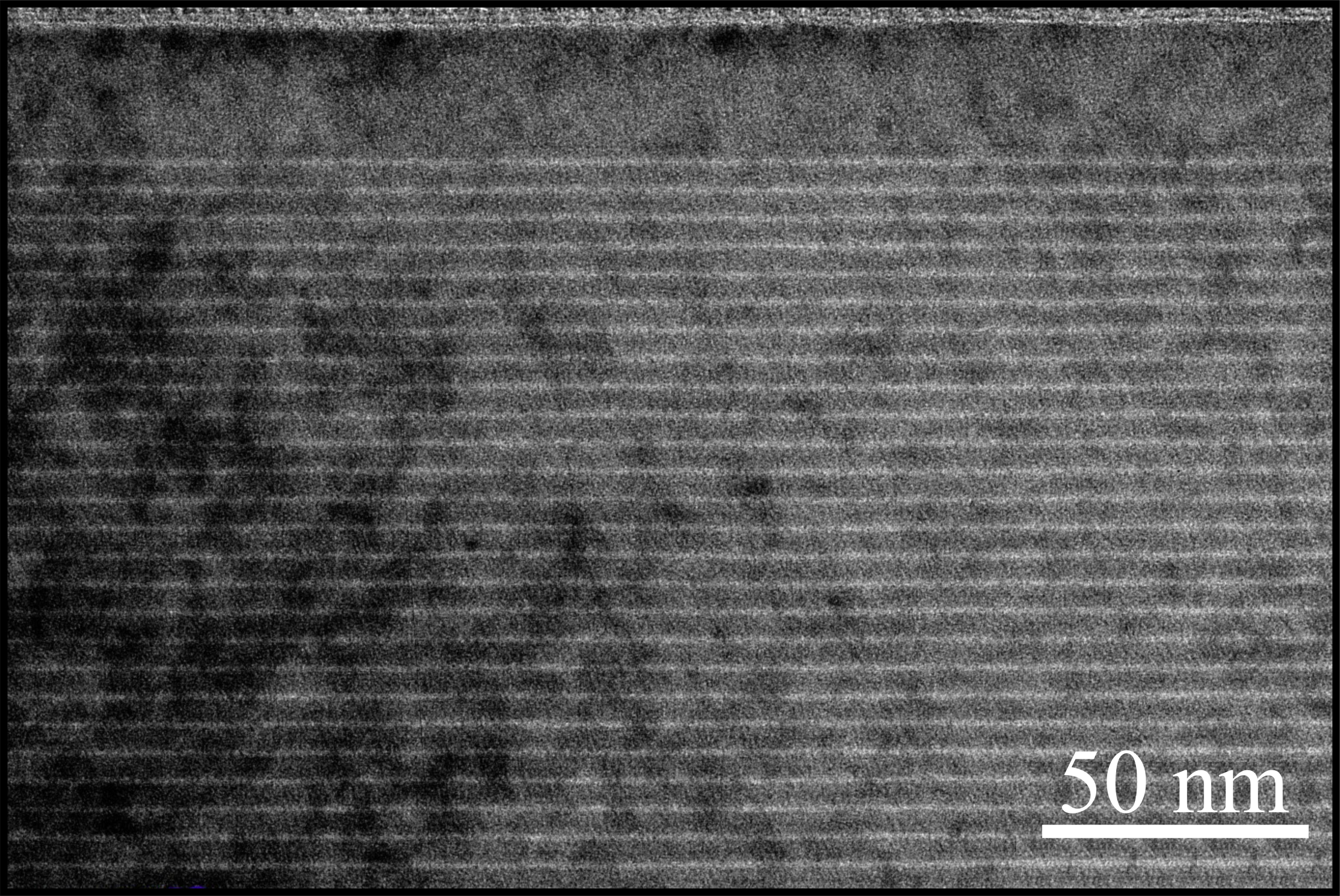}
			}
		\end{center}
		\label{fig:TEM} 
		\caption{TEM images of the superlattice taken at two different scales 
		to justify accuracy of the fabricated structure. Length of 181.94 nm 
		results in 6.06~nm superlattice period. The images are recorded in the 
		microscope Tecnai G2 F20 X-TWIN (FEI Company, Netherlands). Sample for 
		TEM imaging is prepared using FESEM-FIB Helios NanoLab 650 (FEI 
		Company, Netherlands).}
	\end{figure*}
	
	The accuracy of the grown superlattice was checked by TEM and the images 
	are presented in Fig.~\ref{fig:TEM}. The SL processing was done using wet 
	etching to fabricate square mesa of $\rm{80\times80\,\upmu m}$ dimensions 
	and of $\rm{1.3\,\upmu m}$ height. The processed structure was mounted 
	inside the standard single-ridged waveguide; the top contact of the 
	superlattice was attached using $\rm{\sim25\,\upmu m}$ Au wire to ensure 
	suitable coupling with microwave electric fields simultaneously serving for 
	biasing of the structure. The bottom contact was isolated via filter to 
	diminish microwave radiation effect on the coaxial dc-biasing line.
	
\begin{figure*}[h]
		\includegraphics[width=\textwidth]{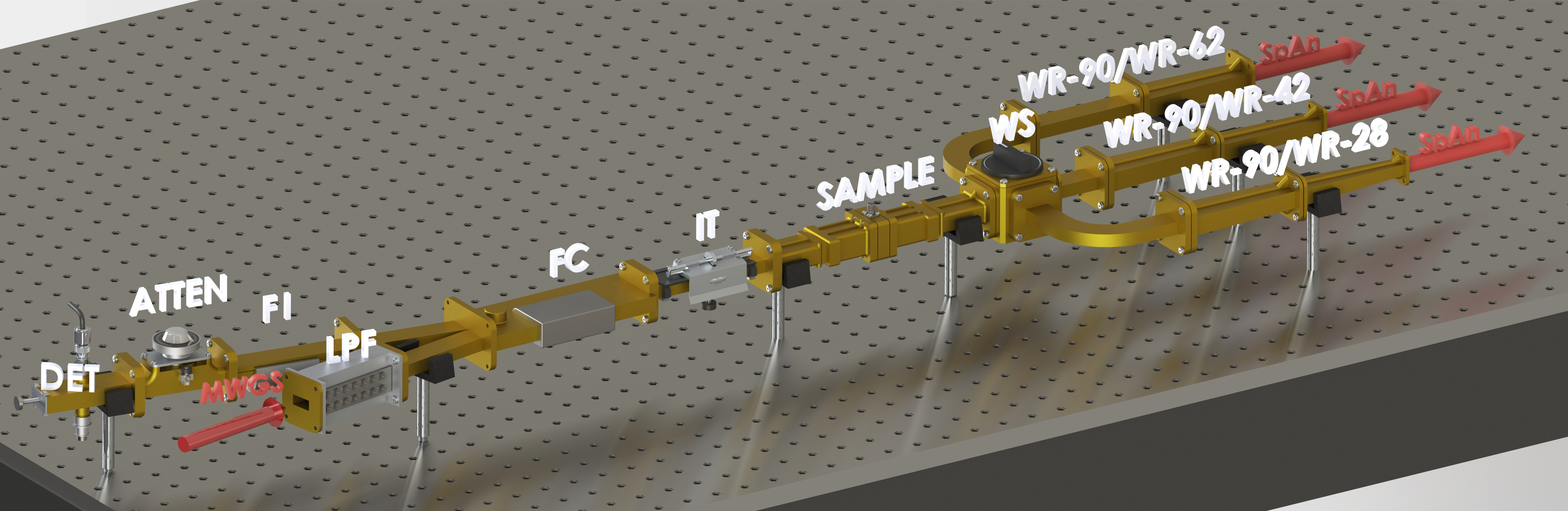}
		\caption{\label{fig:setup} Experimental setup. 
			Microwave radiation generated by klystron (operating within 8.2 - 
			12.4 GHz frequency range), at (MWGS) input is transferred into 
			experimental setup via ferrite circulator (FC). Low pass filter 
			(LPF) is additionally used to remove parasitic high-frequency 
			components. A waveguide switch (WS) was used to connect three 
			WR-90/WR-XX waveguide adapters (WR-XX: WR-62, WR-42, WR-28) to 
			spectrum analyser which allows to perform spectral measurements in 
			a wide frequency range. Impedance transformer (IT) adjusted to 	
			excitation frequency was used to suppress reflections of the 
			excitation signal and ensure maximal signal via proper adjustment 
			of optimal phase at the sample. Reflected wave transient change 
			monitoring was performed at Schottky-diode detector (DET), 
			connected via attenuator (ATTEN), ferrite circulator (FC) and 
			isolator (FI). Sample holder of above-mentioned dimensions was 
			connected via two adapters.}
	\end{figure*}

	Measurements were performed using microwave waveguide-based setup 
	\cite{Subacius2015} relying on the change in transmitted and reflected 
	microwaves induced by the carrier transport processes in the biased SL. The 
	measurement setup is presented in Fig.~\ref{fig:setup}. Microwave radiation 
	of 10~mW power was delivered from a klystron generator operating within 	
	$\rm{8.2-12.4\,GHz}$ frequencies and transferred into the experimental  
	setup via ferrite circulator. To achieve wide frequency range measurements, 
	three different-size waveguide adapters [WR\nobreakdash-90 (X-band) to 	
	WR\nobreakdash-62 $\rm{(K}_u$\nobreakdash-band), WR\nobreakdash-42 	
	(K\nobreakdash-band), and WR\nobreakdash-28 $\rm{(K}_a$-band)] were 
	connected via waveguide switch. An impedance transformer was used to adjust 
	the phase of the microwave radiation in the SL at selected frequency. It 
	was realized by varying the distance between two quartz quarter wave plates 
	inside the transformer enabling hence to achieve maximal response signal 
	from the SL under test.
	
	The reflected signal analysis was monitored using calibrated Schottky diode 
	detector. The frequency response was explored using C4-27 spectrum 
	analyser. All experiments were performed at room temperature. 
	
	\section{NL-criterion for the case of ac-driven superlattice}
	
	In the case of the Esaki-Tsu dependence of the drift velocity on the 
	electric field and for
	\begin{equation}
		\label{sup_NCAEW_eq1}
		NL \leq \frac{7 \epsilon E_{\text{cr}}}{e},
	\end{equation}
	no high-electric field domains are formed in the SL in conditions of NDV 
	(here $N$ is the electron concentration, $L$ is the length of the SL, 	
	$\epsilon=\epsilon_0\epsilon_r$ and $\epsilon_r$ is the relative dielectric 
	constant) \cite{Ignatov1985, Savvidis2004}. The criterium 
	(\ref{sup_NCAEW_eq1}) has been formally derived when only dc bias is 
	applied, but it still can be easily extended to the case of the quasistatic 
	ac electric field [Eq.~(\ref{pump_def})] for which $\omega_0\tau \ll 1$. 
	Indeed, the miniband electrons interact with the quasistatic electric field 
	essentially in the same manner as with the dc electric field, except the 
	total dc+ac electric field is slowly sweeping within the range from 
	$E_{\rm{DC}}-E_0$ to $E_{\rm{DC}}+E_0$ in the course of time. However, the 
	inequality (\ref{sup_NCAEW_eq1}) actually states that the electric 
	stability exists for any applied field strength (see, e.g., Sec. 1.4.1 in 
	\cite{Rieder2004}). Therefore, this form of $NL$-criterion also works if 
	quasistatic electric fields are applied to a quantum superlattice with 
	proper parameters.
	
	\section{Degenerate and nondegenerate amplification of microwaves in 
		superlattices: Calculation of the high frequency mobility within the 
		quasistatic approximation}
	
\begin{figure}[h]
		\includegraphics[width=\textwidth]{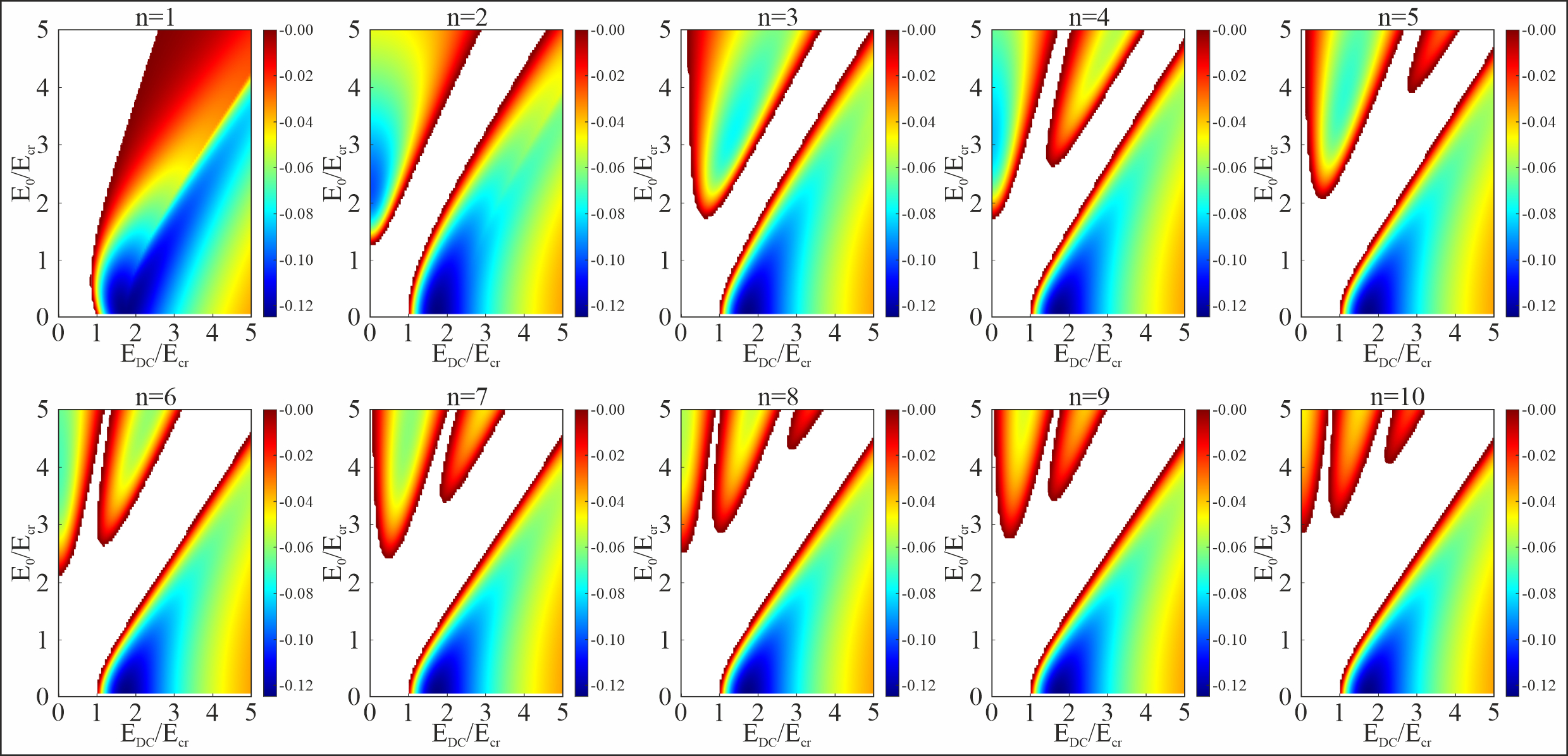} 
		\caption{\label{fig:2Dmaps}Areas and magnitudes of the gain at optimal 	
		phase $\mu_\text{n}^\text{opt}/\mu_0<0$ [Eq.~(\ref{mu-summary2})] in 
		the plane of dc bias $E_{\rm{dc}}$ and pump field amplitude $E_0$ for 
		$n=1:10$. The dc and ac electric fields are scaled to critical field 
		$E_{\rm{cr}}$. Blank areas correspond to absorption ($\mu_n>0$).}
		\label{fig-opt-maps}
	\end{figure}

	Our main aim here is to estimate the high-frequency (HF), small-signal 
	parametric gain in units of the miniband Drude mobility. The major findings 
	are summed up in Fig.~\ref{fig:2Dmaps}.
	
	\subsection{Introduction to the problem}
	
	We assume that the total electric field $E(t)$, which is acting on miniband 
	electrons, consists of the dc and ac pump 
	\begin{equation}
		\label{pump_def}
		E_{\rm{p}}(t)=E_{\rm{DC}}+E_0 \cos(\omega_0 t)
	\end{equation}
	and the ac probe $E_{\rm{pr}}(t)$ fields. We consider two situations for 
	which either 
	\begin{equation}
		\label{nondegenerate1}
		E_{\rm{pr}}(t)=E_1\cos(\omega_1 t+\phi_1)+E_2\cos(\omega_2 t+\phi_2) 
		\quad 
		\rm{with} \quad \omega_1\pm \omega_2=n_\pm\omega_0
	\end{equation}
	or 
	\begin{equation}
		\label{degenerate1}
		E_{\rm{pr}}(t)=E_1\cos(\omega_1 t+\phi) \quad \rm{with} \quad 
		2\omega_1=n_0\omega_0,
	\end{equation}
	where $n_\pm$ are positive integers and $n_0$ is odd; the frequencies 
	$\omega_{1,2}$ in Eq. (\ref{nondegenerate1}) are incommensurable. The cases 
	of Eqs. (\ref{nondegenerate1}) and (\ref{degenerate1}) correspond to the 	
	nondegenerate and degenerate parametric amplifications, respectively. In 
	the both cases all fields must be coherent and therefore the phase 
	differences ($\phi, \phi_{1,2}$) are well-defined constants. In what 
	follows we suppose that the pump $E_{\rm{p}}(t)$ is arbitrary strong, while 
	the probe field is weak ($|E_{\rm{pr}}| \ll |E_{\rm{p}}|$). In contrast to 
	a spontaneous process of frequency multiplication due to nonlinearity 
	$\omega_0\rightarrow n\omega_0$, the parametric amplification is a 
	stimulated processes for which influence of even very weak probe field 
	cannot be ignored. In a single miniband superlattice, the electron drift 
	velocity $v$ depends on the dc electric field as
	\begin{equation}
		\label{esaki-tsu1}
		v(E_{\rm{dc}})=\frac{2v_{\rm{p}}(E_{\rm{dc}}/E_{\rm{cr}})}{[1+(E_{\rm{dc
		}}/E_{\rm{cr}})^2]},
	\end{equation}
	where $v_{\rm{p}}=\Delta d/4\hbar$ is the peak electron velocity (here $d$ 
	is the SL period, $\Delta$ is the miniband width) and $E_{\rm{cr}}=\hbar/e 
	d\tau$ is the Esaki-Tsu critical field ($\tau$ is a characteristic 
	relaxation time). In the quasistatic approximation that is valid for 
	$\omega_0\tau\ll 1$ and $\omega_{1,2}\tau\ll 1$, the drift velocity follows 
	variations of the ac electric fields, and therefore the dependence of $v$ 
	on the total time-dependent field $E(t)$ is given by Eq.~(\ref{esaki-tsu1}) 
	with account of the substitution $E_{\rm{dc}}\rightarrow E(t)$. 
	
	Positive gain is defined in terms of the complex HF mobility 
	$\mu=\mu_{\rm{r}}+i\mu_{\rm{i}}$ as a response of the electron velocity 
	$v(t)$ to the probe ac field:
	\begin{equation}
		\label{mu-def1}
		v=\Re\left[\mu E_{1}e^{-i\omega_1 t}\right]=\left[ 
		\mu_{\rm{r}}\cos(\omega_1 t)+\mu_{\rm{i}}\sin(\omega_1 t) \right] 
		E_{1}, 
		\quad
		\mu_{\rm{r}}=\frac{2}{E_1} \left\langle v(t)\cos(\omega_{1}t) 
		\right\rangle_t, \qquad
		\mu_{\rm{i}}=\frac{2}{E_1} \left\langle v(t)\sin(\omega_{1}t) 
		\right\rangle_t,
	\end{equation}
	where $\left\langle\ldots\right\rangle_t$ denotes averaging over time. In 	
	practical terms, we need to average over a common period of the probe and 
	pump ac fields. In the case of quasistatic fields the imaginary component 
	$\mu_{\rm{i}}$ typically approaches zero [see, e.~g.,  
	Eq.~(\ref{sin-comonent1})], and thus we will focus on the behaviour of 	
	$\mu\equiv\mu_{\rm{r}}$ as a function of the pump field parameters. It is 	
	natural to scale $\mu$ to the Drude mobility 	
	$\mu_0=2v_{\rm{p}}/E_{\rm{cr}}=e\tau/m_0$ ($m_0=\frac{2\hbar^2}{\Delta 
	d^2}$ is the effective electron mass at the bottom of the miniband). Note 
	that the Drude mobility is related to the Drude conductivity $\sigma_0$ as 
	$\mu_0=\sigma_0/(eN)$.
	
	\subsection{Summary of the main results on HF mobilities}
	
	We found that for the both types of parametric amplification 
	[Eqs.~(\ref{nondegenerate1}, \ref{degenerate1})] the small-signal HF 
	mobility $\mu_{\rm{n}}(\psi)$ can be represented as the following sum of 
	the phase-independent \textit{incoherent HF mobility} $\mu_{\rm{inc}}$ and 
	the phase-dependent \textit{coherent HF mobility} $\mu_{\rm{coh}}(\psi)$:
	\begin{equation}
		\label{mu-summary1}
		\mu_{\rm{n}}(\psi)=\mu_{\rm{inc}}+\mu_{\rm{coh}}(\psi), \quad
		\mu_{\rm{inc}}=\mu_0\left 
		\langle\frac{(1-F_{\rm{p}}^2)}{(1+F_{\rm{p}}^2)^2} \right \rangle_T, 
		\quad
		\mu_{\text{coh}}(\psi)=\mu_0\left 
		\langle\frac{(1-F_\text{p}^2)}{(1+F_{\rm{p}}^2)^2}\cos(n\omega t) 
		\right 
		\rangle_T\cos\psi,
	\end{equation}
	where $F_\text{p}(t)=E_\text{p}(t)/E_{\text{cr}}$ is the scaled pump field 
	[Eq.~(\ref{pump_def})], the sign $\left\langle\ldots\right\rangle_T$ 
	denotes an average $\frac{1}{T}\int_0^{T}dt$ over the period of the pump 
	field $T=2\pi/\omega$, and two other parameters (relative phase and 
	parametric order) are
	\begin{equation}
		(\psi;  n)=   \left\{ 
		\begin{array}{cccc}
			( \phi_1+\phi_2; & n_{+} ),  & \mbox{ see 
			Eq.~(\ref{nondegenerate1}) } 
			&  \mbox{ (nondegenerate sum), } \\ 
			( \phi_1-\phi_2; & n_{-} ),  & \mbox{ see 
			Eq.~(\ref{nondegenerate1}) } 
			& \mbox{ (nondegenerate difference), } \\   
			( \phi;          & n_0  ),   & \mbox{ see Eq.~(\ref{degenerate1}) }
			& \mbox{ (degenerate). }
		\end{array}
		\right]
	\end{equation}
	The coherent gain reaches its maximum at an optimal phase 
	$\psi=\psi_{\text{opt}}$, and the corresponding value of the total mobility 
	$\mu_\text{n}(\psi_{\text{opt}})$ is
	\begin{equation}
		\label{mu-summary2}
		\mu^{\text{(opt)}}_n=\frac{\mu_0}{2\pi}\int_0^{2\pi}\frac{(1-F^2_\text{p
		})}{(1+F_\text{p}^2)^2}dx - \frac{\mu_0}{2\pi}\left | 
		\int_0^{2\pi}\frac{(1-F_\text{p}^2)\cos(nx)}{(1+F_\text{p}^2)^2}dx 
		\right |,
	\end{equation}
	where $F_\text{p}(x)=F_{\text{DC}}+F_\text{0}\cos(x)$. 
	Note that the integrals in Eq.~(\ref{mu-summary2}) can be taken 
	analytically and result in cumbersome functions of complex variables [see 
	Eq.~(9) in \cite{Romanova2012}]. We underline that $\mu^{\text{(opt)}}_n$ 
	depends on the electric fields 
	${F_{\text{DC}}=E_{\text{DC}}/E_{\text{cr}}}$ and 	
	$F_0=E_0/E_{\text{cr}}$, but not on the pump frequency $\omega_0$ itself.
	Figure~\ref{fig-opt-maps} shows the areas of optimal gain 
	$\mu^{\text{(opt)}}_n<0$ together with its magnitude in units of the Drude 
	mobility $\mu_0$ for different numbers of the pump photons $n$. This figure 
	provides an extension of Fig. 3(a-b) of the main text for $n>2$. Note that 
	relative positions of the gain and absorption areas for $n>2$ are similar 
	to those in the case $n=2$. 
	
	In the coming subsections we will derive the 
	Eqs.~(\ref{mu-summary1}-\ref{mu-summary2}) for both the degenerate and 
	tnondegenerate schemes of amplification following the quasistatic approach 
	of our earlier works \cite{Shorokhov2006, Alekseev2006}.
	
	\subsection{Calculations of HF mobilities for the degenerate case}
	
	In the case of degenerate amplification, the HF mobility $\mu_{n_0}(\phi)$ 
	[cf. Eq.~(\ref{mu-def1})] scaled to the Drude mobility $\mu_0$ can be found 
	as 
	\begin{equation}
		\label{A-degen1}
		\frac{\mu_{n_0}(\phi)}{\mu_0}=\frac{E_{\text{cr}}A(\phi)}{v_p E_1}, 
		\quad 
		A(\phi)=\langle v[E(t)]\cos(\omega_{1}t+\phi)\rangle_t,
	\end{equation}
	where $E(t)=E_\text{p}(t)+E_{\text{pr}}(t)$ and the pump and probe electric 
	fields are given by Eqs.~(\ref{pump_def},\ref{degenerate1}), respectively. 
	Following Eq.~(\ref{A-degen1}) we essentially need to calculate $A(\phi)$, 
	and the amplification corresponds to $A<0$. Since the probe electric field 
	is weak, we can use the truncated Taylor series for the drift velocity
	\begin{equation}
		\label{taylor}
		v(E)\approx v[E_{\text{p}}(t)]+v^\prime[E_\text{p}(t)] \, 
		E_1\cos(\omega_{1}t+\phi), \quad
		v^\prime[E_\text{p}(t)] \equiv \frac{\partial v}{\partial E}\biggl| 
		_{\text{E=E}_\text{p}\text{(t)}}
	\end{equation}
	and then substitute it in Eq. (\ref{A-degen1}). This results in 
	\begin{equation}
		\label{A-degen2}
		A=A_{\text{h}}+A_{\text{inc}}+A_{\text{coh}}=\left\langle 
		v[E_\text{p}(t)]\cos(\omega_{1}t+\phi) \right\rangle_{\text{2T}}+
		(E_1/2) \left\langle v^\prime[E_\text{p}(t)] \right\rangle_{\text{2T}} +
		(E_1/2) \left\langle v^\prime[E_\text{p}(t)]\cos(2\omega_{1}t+2\phi) 
		\right\rangle_{\text{2T}}. 
	\end{equation}
	For the electron velocity Eq. (\ref{esaki-tsu1}) and its derivative
	\begin{equation}
		\label{esaki-tsu2}
		\frac{\partial v}{\partial 
			E}=\left(\frac{2v_\text{p}}{E_{\text{cr}}}\right) 
		\frac{1-(E/E_{\text{Cr}})^2}{(1+(E/E_{\text{Cr}})^2)^{2 }},
	\end{equation}
	the Eq. (\ref{A-degen2}) can be further simplified to
	\begin{equation}
		\label{A_cohA_incoh1}
		A_{\text{inc}}= \left\langle 
		v^\prime[E_\text{p}(t)]\right\rangle_{\text{T}}\cdot (E_1/2), \quad 
		A_{\text{coh}}(\phi)=\left\langle v^\prime[E_\text{p}(t)]\cos(n_0\omega 
		t) \right\rangle_{\text{T}} \, \cos(2\phi)\cdot (E_1/2)
	\end{equation}
	and $A_{\text{h}}=\cos\phi \, \left\langle v[E_p(t)]\cos(\omega_{1}t) 
	\right\rangle_{\text{2T}}=0$. Now we substitute the derivative 
	Eq.~(\ref{esaki-tsu2}) in Eqs.~(\ref{A_cohA_incoh1}) and then, following 
	the findings of Eq.~(\ref{A-degen1}), divide the resulting equations by 
	$(v_\text{p} E_1/E_{\text{Cr}})$. In such a way we obtain the desirable 
	form of HF mobility $\mu_{\text{n}_\text{0}}(\phi$) [see 
	Eq.~(\ref{mu-summary1})].
	
\begin{figure}
		\includegraphics[width=0.6 \columnwidth]{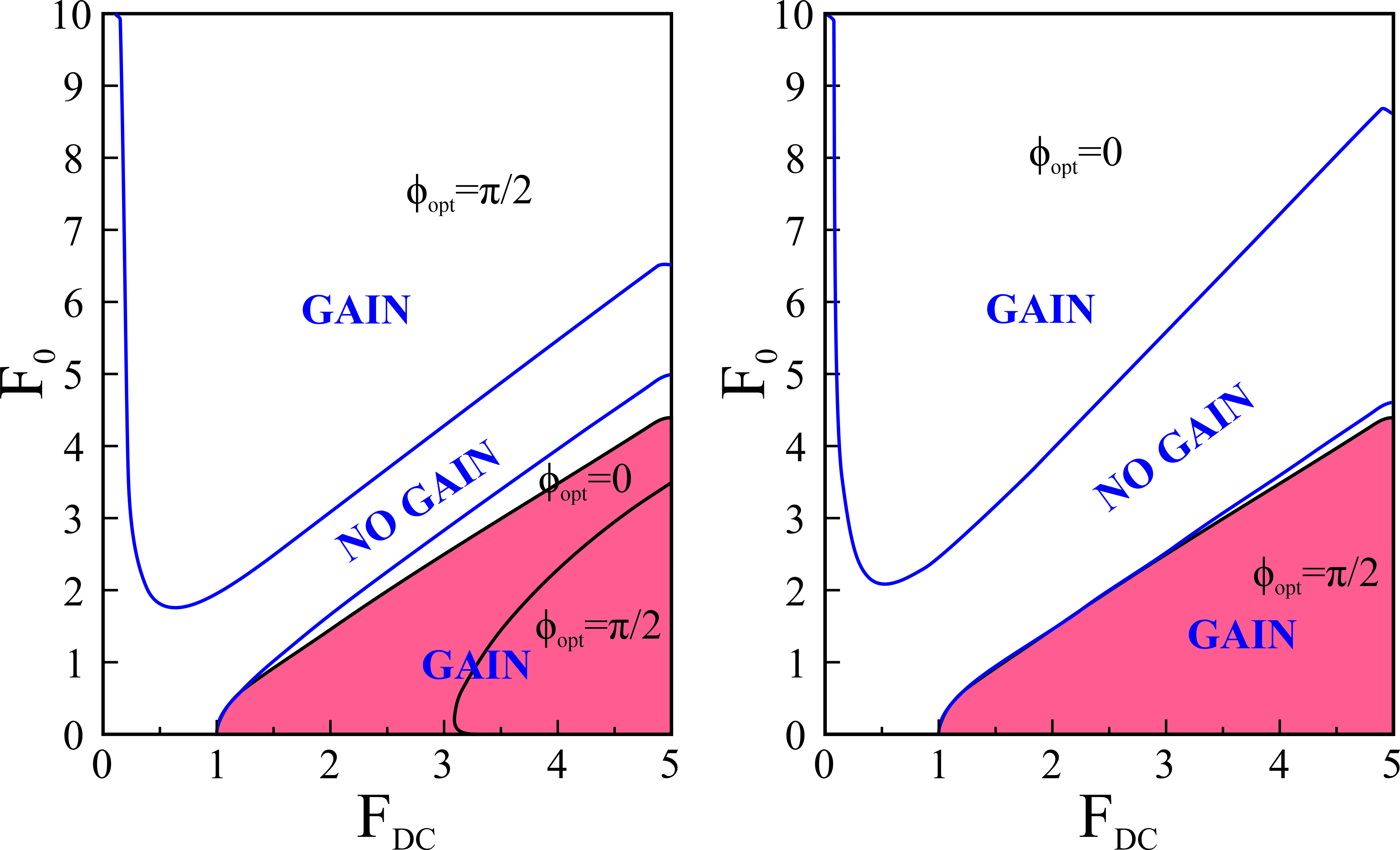} 
		\caption{Regions of the net small-signal gain 
			$A=A_{\text{inc}}+A_{\text{coh}}<0$ [Eq.~(\ref{A-degen2})] and the 
			corresponding optimal phases $\phi_{\text{opt}}$ 
			[Eq.~(\ref{phi-opt1})] in the $F_{\text{DC}-F_0}$ plane. Pink marks 
			regions of the incoherent gain $A_{\text{inc}}<0$ 
			[Eq.~(\ref{A_cohA_incoh1})]. Left panel: $n_0=3$, right 			
			panel: $n_0=5$.}
		\label{fig-opt-phase}
	\end{figure}

	The phase-dependant gain term $A_{\text{coh}}(\phi)<0$ has its maximal 
	value ($\max|A_{\text{coh}}|$) at some optimal phase $\phi_{\text{opt}}$ 
	that can be determined from the condition $\frac{\partial 
	A_{\text{coh}}}{\partial\phi}=0$ 
	as 
	\begin{equation}
		\label{phi-opt1}
		\left\langle v^\prime[E_\text{p}(t)]\cos(2\omega_{1}t) 
		\right\rangle_\text{T} \,\sin(2\phi_{\text{opt}})=0, \quad
		\phi_{\text{opt}}=:  0, \pi/2.
	\end{equation}
	The exact value that $\phi_{\text{opt}}$ takes for the given dc bias 
	$E_{\text{DC}}$ and the amplitude of ac pump $E_0$ [Eq.~(\ref{pump_def})] 
	depends on the sign of the integral involved in Eq.~(\ref{phi-opt1}), as 
	illustrated in Fig.~\ref{fig-opt-phase} for the cases 
	$\omega_{1}/\omega=3/2$ and $5/2$. Nevertheless, whatever is exact value of 
	the optimal phase, $A_{\text{coh}}(\phi_{\text{opt}})$ still has the form 
	\begin{equation}
		\label{phi-opt2}
		A_{\text{coh}}(\phi_{\text{opt}})=-\Bigl| \left\langle 
		v^\prime[E_\text{p}(t)]\cos(n_0\omega t) \right\rangle_\text{T} \Bigr| 
		\cdot (E_1/2).
	\end{equation}
	
	\subsection{Calculations of HF mobilities for the nondegenerate cases}
	
	To calculate gain in two nondegenerate cases $\omega_1\pm 
	\omega_2=n_\pm\omega$ [Eq.~(\ref{nondegenerate1})] we essentially need to 
	find (negative) values of the following integrals 
	\begin{equation}
		\label{A-nondegen1}
		A_{\omega1}= \left\langle v[E(t)]\cos(\omega_1 t+\phi_1)\right 
		\rangle_t, 
		\quad
		A_{\omega2}=\langle v[E(t)]\cos(\omega_2 t+\phi_2)\rangle_t,
	\end{equation}
	where $E(t)=E_\text{p}(t)+E_{\text{pr}}(t)$ and $E_\text{p}(t)$, 
	$E_{\text{pr}}(t)$ are defined by 
	Eqs.~(\ref{pump_def},\ref{nondegenerate1}). Now $E(t)$ is a quasiperiodic 
	function of time, and therefore averaging in Eqs. (\ref{A-nondegen1}) is 
	performed over infinite time. Since the expressions of $A_{\omega1}$ and 
	$A_{\omega2}$ are similar, we will only focus on the calculations related 
	to $A_{\omega1}$. For $E_{1,2}\ll E_{\text{0,DC}}$ we make the expansion 
	similar to Eq.~(\ref{taylor}) and obtain 
	\begin{equation}
		A_{\omega1}= \left\langle v[E_\text{p}(t)]\cos(\omega_1 t+\phi_1) 
		\right\rangle_\text{t} + \left\langle \frac{\partial v}{\partial 
		E}\biggl|_{\text{E}_\text{p(t)}} \{ E_1\cos(\omega_1 
		t+\phi_1)+E_2\cos(\omega_2 
		t+\phi_2) \} \cdot \cos(\omega_1 t+\phi_1) \right\rangle_\text{t} =
		\nonumber
	\end{equation}
	\begin{equation}
		\label{A-nondegen2}
		= \frac{E_1}{2} \left\langle \frac{\partial v}{\partial E}\biggl| 
		_{\text{E}_\text{p(t)}} \right\rangle_\text{t} + \frac{E_2}{2} 
		\left\langle 
		\frac{\partial v}{\partial E}\biggl| _{\text{E}_\text{p}} \cos(\omega_1 
		t+\omega_2 t) \right\rangle_\text{t} \, \cos(\phi_1+\phi_2)+ 
		\frac{E_2}{2} \left\langle \frac{\partial v}{\partial E}\biggl| 
		_{\text{E}_\text{p}} \cos(\omega_1 t-\omega_2 t) \right\rangle_\text{t} 
		\, \cos(\phi_1-\phi_2).
	\end{equation}
	By assuming that $\omega_1+\omega_2=n_{+}\omega$ and 
	$\omega_1-\omega_2=n_{-}\omega$ are satisfied, we can represent Eq. 
	(\ref{A-nondegen2}) in the form 
	\begin{equation}
		A^{(\omega1)}=A^{(\omega1)}_{\text{inc}}+A^{(\omega1)}_{\text{coh}}(\phi
		_{+} )+A^{(\omega1)}_{\text{coh}}(\phi_{-}),
		\nonumber
	\end{equation}
	\begin{equation}
		\label{A-nondegen3}
		A^{(\omega1)}_{\text{inc}}= \left\langle 
		v^\prime[E_\text{p(t)}]\right\rangle_\text{T}\cdot (E_1/2), \quad
		A^{(\omega1)}_{\text{coh}}(\phi_\pm)= \left\langle 
		v^\prime[E_\text{p(t)}]\cos(n_\pm\omega t)\right\rangle_\text{T} \, 
		\cos(\phi_\pm)\cdot (E_2/2),
	\end{equation}
	where $\phi_\pm\equiv\phi_1 \pm \phi_2$ and the derivative $v^\prime(E)$ is 
	given by Eq. (\ref{esaki-tsu2}). The optimal values of the phase $\phi_{+}$ 
	(or $\phi_{-}$) are $\phi^{\text{opt}}_{+}=: 0,\, \pi$ [cf. Eq. 
	(\ref{phi-opt1})].
	
	We see that the integrals involved in the incoherent terms $A_{\text{inc}}$ 
	for degenerate and nondegenerate cases are identical, while the 
	corresponding coherent terms $A_{\text{coh}}$ coincide when $n_\pm$ are odd 
	numbers [Eqs.~(\ref{A_cohA_incoh1}) vs. Eqs.~(\ref{A-nondegen3})]. 
	Furthermore, for even values of $n_\pm$ the mathematical structure of 
	$A_{\text{coh}}$ terms is still the same, and this allows us to represent 
	all small-signal mobilities, derived within the quasistatic limit, as the 
	universal expressions Eq. (\ref{mu-summary1}).
	
	Finally, we want to demonstrate that the imaginary part of the HF mobility 
	$\sim\left\langle v[E(t)]\sin(\omega_{1}t+\phi_1)\right\rangle_\text{t}$ 
	[cf. Eq.~(\ref{mu-def1})] vanishes. For this aim we again use the Taylor 
	expansion for $v(E)$, substitute it in $\left\langle 	
	v[E(t)]\sin(\omega_{1}t+\phi_1)\right\rangle_\text{t}$, and obtain 
	\begin{equation}
		\left\langle v[E_\text{p}(t)]\sin(\omega_1 t+\phi_1) 
		\right\rangle_\text{t} + \left\langle \frac{\partial v}{\partial 
		E}\biggl| _{\text{E}_\text{p(t)}} 
		\left\{ E_1\cos(\omega_1 t+\phi_1)+E_2\cos(\omega_2 t+\phi_2) \right\} 
		\cdot \sin(\omega_1 t+\phi_1) \right\rangle_\text{t} =
		\nonumber
	\end{equation}
	\begin{equation}
		\label{sin-comonent1}
		= \frac{E_2}{2} \left\langle \frac{\partial v}{\partial E}\biggl| 
		_{\text{E}_\text{p}} \cos(n_{+}\omega t) \right\rangle_\text{T} \, 
		\sin\phi_{+}- \frac{E_2}{2} \left\langle \frac{\partial v}{\partial 
		E}\biggl| _{\text{E}_\text{p}} \cos(n_{-}\omega t) 
		\right\rangle_\text{T} 
		\,\sin\phi_{-},
	\end{equation}
	which becomes zero when $\phi_\pm\rightarrow\phi^{\text{opt}}_\pm$.
	
	\section{Estimation of the pump electric field strength in the SL}
	\label{sup_estimation}
	
	Taking into account the impossibility to determine strength of the pump 
	field inside the SL using current experimental setup, estimation, based on 
	matching experimentally achieved emission power dependency on biasing 
	voltage and above-described theoretical model, was performed in order to 
	find at least approximate working point position.
	
	Fitting the theoretical results given in Fig.~\ref{fig:2Dmaps} to 
	experimental data one needs to note that several multiphoton processes 
	undergo on the same time. Consequently, one can expect that the total 
	mobility is a result of superposition of different multiphoton processes. 
	Theoretical model, describing such superposition is an object of future 
	investigations, but for the spectra, presented in Fig.~2 of the main body, 
	these relations can be determined using emission power dependency on 
	biasing voltage. Simple linear dependency was used as a point of departure:
	
	\begin{equation}
		\left( \frac{\mu }{\mu _{0}} \right)_{\text{total}} = \sum C_{\rm{n}} 
		\cdot \frac{\mu_{\rm{n}} }{\mu_0}
	\end{equation}
	
	where \textit{n} is integer and $C_\text{n}$ is weight coefficient. It is 
	reasonable to assume that $C_\text{n}$ will be inversely proportional to 
	\textit{n} indicating that mobilities with small \textit{n} will give the 
	essential influence for the final mobility dependence.
	
\begin{figure}[h]
		\begin{center}
			\resizebox{\textwidth}{!}{
				\includegraphics{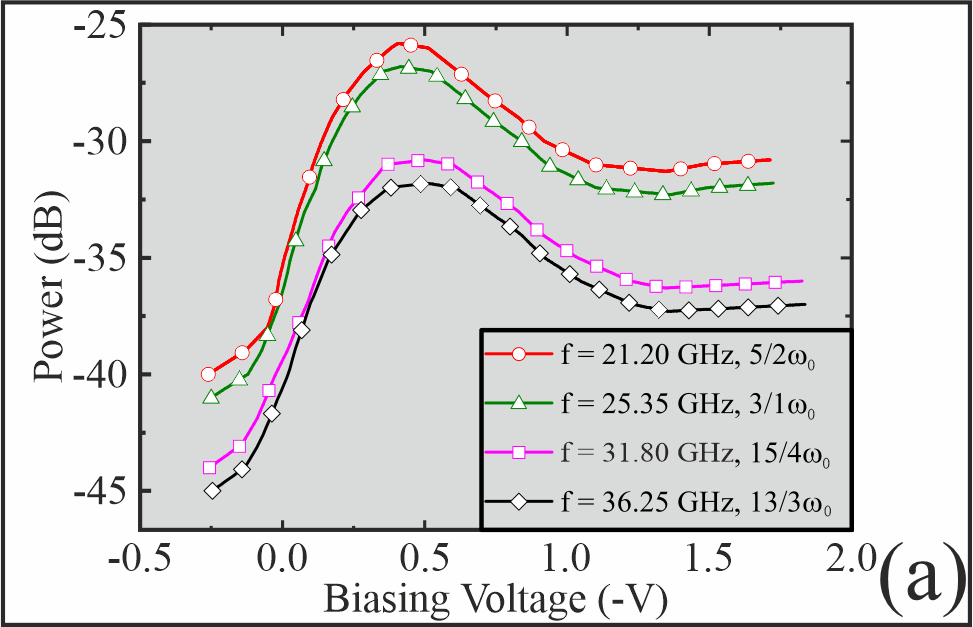}
				\includegraphics{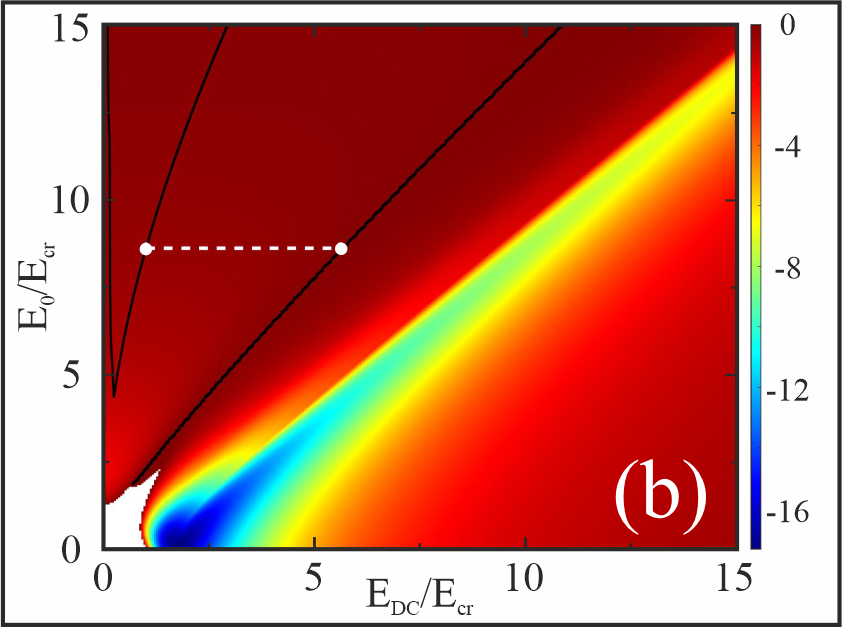}
				\includegraphics{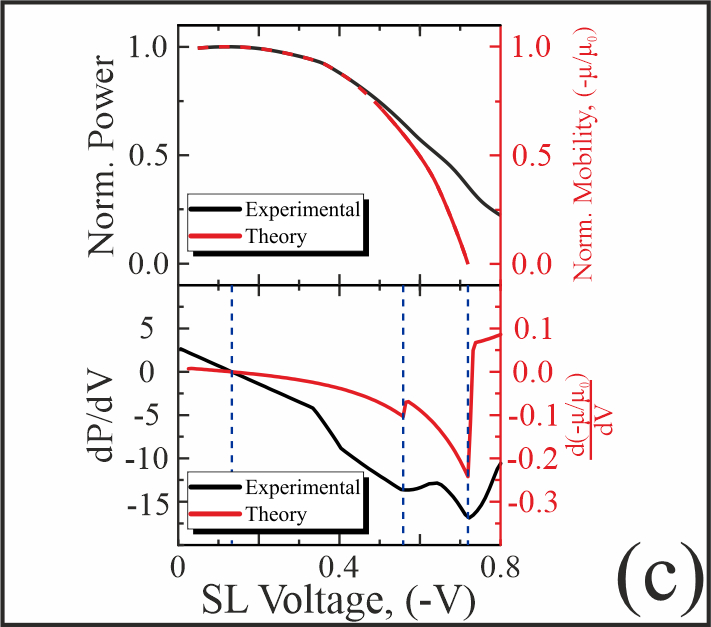}
			}
		\end{center}
		\caption{(a): Emission power dependence on the SL bias voltage at 
		different frequencies acquired using calibrated reference. Note 
		equivalent evolution of all the spectral constituents the with DC bias.
		(b): Theoretical sketch of superpositioned mobility dependency on 
		DC-biasing and microwave radiation field strength for $5/2 \omega_0$ 
		frequency. Black lines represent mobility peak positions while white 
		dotted line corresponds to the position of matched experimental profile.
		(c): superpositioned $\mu_{\rm{total}}(\phi_{\rm{opt}})/\mu_0$ profile 
		and its derivative for $5/2\omega_0$ frequency for the estimated pump 
		field strength ($F_0=8.61$) in SL along with experimental results 
		(built-in voltage included). Analysing derivative lines one can note 
		that theoretical peak positions correspond for experimental data (see 
		blue dashed lines).}
		\label{fig:sup_E_estimate}
	\end{figure}

	In order to get weight coefficients, emission power and mobility ratio 
	dependence on biasing voltage profile equality was assumed. Arbitrary 
	frequency of 21.20~GHz, corresponding to $\frac{5}{2} \omega_0$ was 
	selected and the emitted power dependency on biasing voltage along with its 
	derivative were achieved [see Fig.~\ref{fig:sup_E_estimate}(a)]. For 
	simplicity relations were found including only $n=1,2,3$. Influence of 
	bigger \textit{n} can be assumed as negligibly small. The final 
	superposition relation was found to be as
	
	\begin{equation}
		\left( \frac{\mu }{\mu _{0}} \right)_{\text{total}} = 119 \cdot 
		\frac{\mu_{\rm{n=1}}}{\mu_0} + 18.9 \cdot \frac{\mu_{\rm{n=2}} }{\mu_0} 
		+ 1 \cdot \frac{\mu_{\rm{n=3}}}{\mu_0}
	\end{equation}
	
	and results in mobility distribution, depicted in 
	Fig.~\ref{fig:sup_E_estimate}(b). 
	
	One needs essentially to underline that experimental local derivative 
	minima arise at the same voltage values as ones, obtained from theoretical 
	model [see Fig.~\ref{fig:sup_E_estimate}(c)]. Note that due to 
	superposition no-gain region vanished comparing to the situation when the 
	single \textit{n} mobility distribution is analysed. Also, this calculation 
	results in $F_0 = 8.66$ alternating electric field strength across the 
	superlattice.
	
	\section{The measured input-output power dependencies}
	
	The aim here is to present the emission power dependencies on the pump 
	power (input-output plot) for  subharmonics, arising due to multiphoton 
	parametric processes in the SL, in comparison with the inherent process of 
	frequency doubling.
	
	The method of input-output characteristics is rather often used in 
	nonlinear optics to analyse frequency multiplication and wave mixing 
	phenomena. This indirect method is essentially based on comparison of  
	experimental curves with the corresponding predictions of theoretical 
	models. In the simplest case, the power at the frequency corresponding to 
	$n$-th harmonic is featured by $n^{th}$ power function dependence on the 
	excitation power. Quadratic dependence was observed in experiments on 
	optical second harmonic generation (see, e.~g., \cite{Wang2014} for 
	in-cavity and \cite{Zhao2021} for cavityless SHG measurements). However, 
	the input-output dependence typically becomes linear in the case of 
	conventional optical parametric generation in a medium with a quadratic 
	nonlinearity (cf. \cite{Byer1997} for the theory and \cite{Du2002} for the 
	experiment).
	
\begin{figure}[!h]
		\centering
		\includegraphics[width=0.5\textwidth]{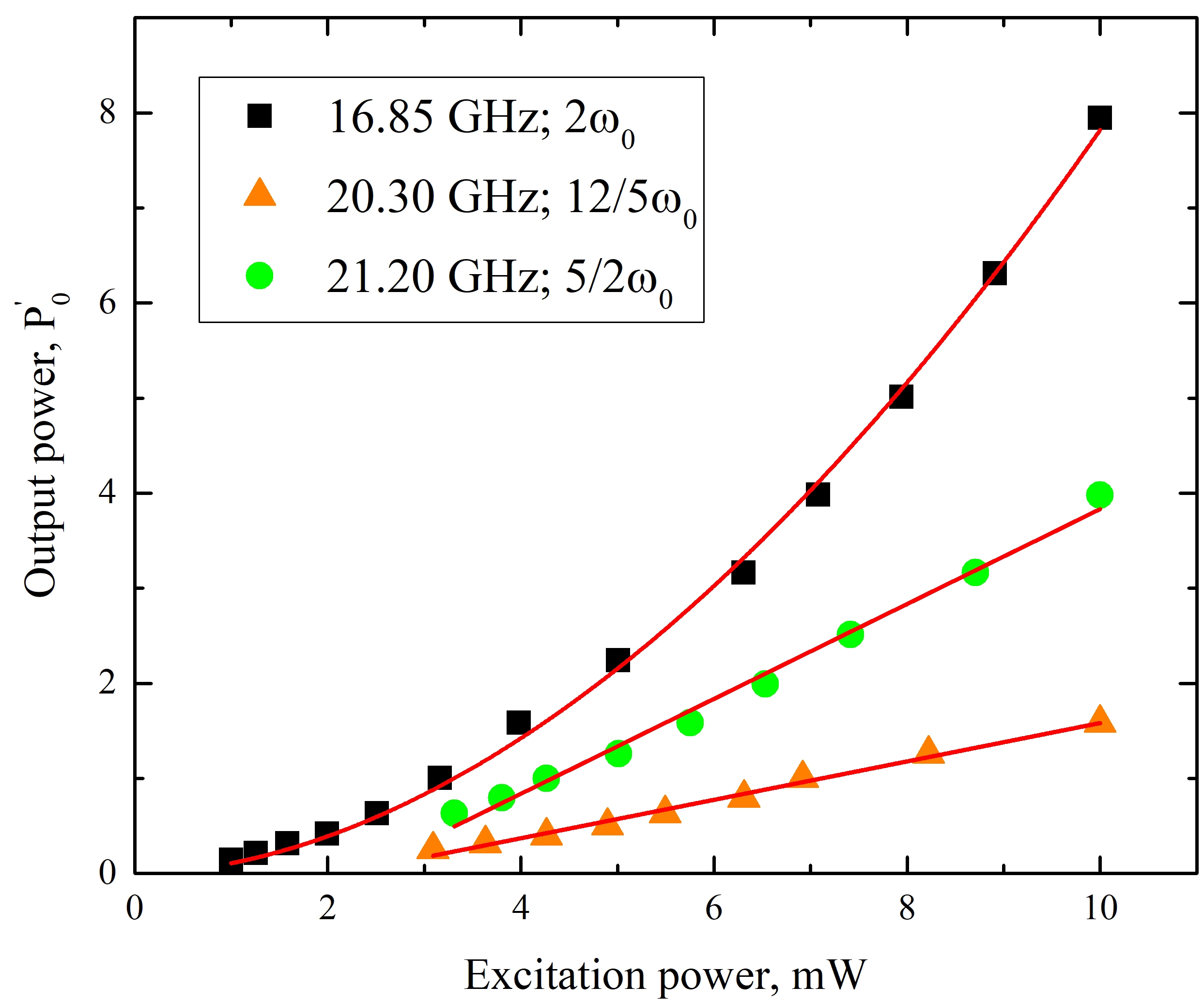}
		\caption{\label{fig:power} Input-output plots for the superlattice 
		device demonstrating dissipative parametric gain. There is no 
		external voltage bias; the pump frequency is 8.45 GHz. Notice that the 
		corresponding dependencies for degenerate and nondegenerate multiphoton 
		parametric processes can be approximated by the straight lines with 
		different gradients.}
	\end{figure}

	We emphasize that the superlattice in microwave electric field is not 
	purely optical system -- it is nonlinear optoelectronic system operating in 
	the negative differential velocity regime with predominating multiphoton 
	processes. The measured input-output dependencies for the studied 
	superlattice are given in Fig.~\ref{fig:power} for three characteristic 
	processes: SHG, 5/2-harmonic generation and fractional ($12/5$) harmonic 
	generation. In the small-signal limit (small output power), the later 
	frequency results from a mixture of 6-photon down-conversion and 3-photon 
	up-conversion processes [fig.~2 in the main body of the paper]. As one can 
	see in Fig.~\ref{fig:power}, the input-output dependence of the second 
	harmonic is seeking the square law, while the corresponding behaviour of 
	both subharmonics is linear but with different slopes. In contrast to its 
	optical counterpart \cite{Byer1997}, the existing theoretical model of 
	dissipative parametric gain in the SL is still not in a mature state to 
	predict the shape of the generated power at a given frequency as a function 
	of the pump strength. Therefore, the input-output method is currently less 
	informative in analysis of the dissipative gain than the direct 
	spectroscopic approach.
	
	In summary, the measured  difference in dependencies between the input and 
	output powers together with the spectroscopic analysis provide convincing 
	experimental evidence of the multiphoton parametric processes.

	\section{Figure 2 data in the form of two tables}
	
	The tables below represent data used in Fig.~2 of the main text. All 
	frequencies are in GHz; the frequency ratios (fractions) are defined with 
	respect to the pump frequency $\omega_0/2\pi=8.45$ GHz.
	
	\def\arraystretch{1.5}
	\begin{table}[H]
		\centering
		\caption{Expanded relations of multiphoton and multiple wave mixing 
		phenomena, related to Fig. 2 main panel.}
		\begin{tabular}{|r|l|c|r|l|c|r|c|l|}
			\hline
			\textbf{Frequency} & \textbf{Fraction} & ~ & \textbf{Frequency} & 
			\textbf{Fraction} & ~ & \textbf{$n_\pm$} & ~ & 
			\textbf{\begin{tabular}[c]{@{}c@{}}Excitation\\ 
			frequency\end{tabular}} \\ 
			\hline
			11.20 & (4/3)  & + &  23.05 & (8/3)  & = &  4   & $\times$ & 8.45 
			\\ \hline
			12.67 & (3/2)  & + &  21.20 & (5/2)  & = &  4   & $\times$ & 8.45 
			\\ \hline
			14.46 & (7/4)  & + &  18.98 & (9/4)  & = &  4   & $\times$ & 8.45 
			\\ \hline
			13.45 & (8/5)  & + &  20.30 & (12/5) & = &  4   & $\times$ & 8.45 
			\\ \hline
			18.98 & (9/4)  & + &  23.05 & (11/4) & = &  5   & $\times$ & 8.45 
			\\ \hline
			13.45 & (8/5)  & + &  28.20 & (17/5) & = &  5   & $\times$ & 8.45 
			\\ \hline
			23.05 & (8/3)  & + &  28.20 & (10/3) & = &  6   & $\times$ & 8.45 
			\\ \hline
			14.46 & (7/4)  & + &  36.25 & (17/4) & = &  6   & $\times$ & 8.45 
			\\ \hline
			18.98 & (9/4)  & + &  31.80 & (15/4) & = &  6   & $\times$ & 8.45 
			\\ \hline
			18.98 & (11/5) & + &  31.80 & (19/5) & = &  6   & $\times$ & 8.45 
			\\ \hline
			23.05 & (8/3)  & + &  36.25 & (13/3) & = &  7   & $\times$ & 8.45 
			\\ \hline
			23.05 & (11/4) & + &  36.25 & (17/4) & = &  7   & $\times$ & 8.45 
			\\ \hline
			31.80 & (15/4) & + &  36.25 & (17/4) & = &  8   & $\times$ & 8.45 
			\\ \hline
			31.80 & (19/5) & + &  36.25 & (21/5) & = &  8   & $\times$ & 8.45 
			\\ \hline
			\multicolumn{9}{c} ~ \\ \hline
			36.25 & (13/3) & - &  28.20 & (10/3) & = &  1   & $\times$ & 8.45 
			\\ \hline
			21.20 & (5/2)  & - &  12.67 & (3/2)  & = &  1   & $\times$ & 8.45 
			\\ \hline
			31.80 & (15/4) & - &  23.05 & (11/4) & = &  1   & $\times$ & 8.45 
			\\ \hline
			23.05 & (11/4) & - &  14.46 & (7/4)  & = &  1   & $\times$ & 8.45 
			\\ \hline
			28.20 & (17/5) & - &  20.30 & (12/5) & = &  1   & $\times$ & 8.45 
			\\ \hline
			28.20 & (10/3) & - &  11.20 & (4/3)  & = &  2   & $\times$ & 8.45 
			\\ \hline
			36.25 & (17/4) & - &  18.98 & (9/4)  & = &  2   & $\times$ & 8.45 
			\\ \hline
			36.25 & (21/5) & - &  18.98 & (11/5) & = &  2   & $\times$ & 8.45 
			\\ \hline
			31.80 & (15/4) & - &  14.46 & (7/4)  & = &  2   & $\times$ & 8.45 
			\\ \hline
			36.25 & (13/3) & - &  11.20 & (4/3)  & = &  3   & $\times$ & 8.45 
			\\ \hline
		\end{tabular}
	\end{table}

	\begin{table}[H]
		\centering
		\caption{Expanded classification of the generated frequencies according 
		to the pump fractions (see inset of Fig. 2).}
		\begin{tabular}{|c|c|c|c|c|}
			\hline
			{\textbf{\begin{tabular}[c]{@{}c@{}}Line \\ number\end{tabular}}} & 
			{\textbf{Frequency}} & \multicolumn{3}{c|}{\textbf{2.5\% accuracy 
			fractions}} \\ \hline
			1  & 8.45  & 1/1 (0.00\%)  & ~             & ~              \\ 
			\hline
			2  & 11.20 & 4/3 (0.59\%)  & ~             & ~              \\ 
			\hline
			3  & 12.67 & 3/2 (0.04\%)  & ~             & ~              \\ 
			\hline
			4  & 13.45 & 8/5 (0.52\%)  & ~             & ~              \\ 
			\hline
			5  & 14.46 & 7/4 (2.21\%)  & ~             & ~              \\ 
			\hline
			6  & 16.85 & 2/1 (0.30\%)  & ~             & ~              \\ 
			\hline
			7  & 18.98 & 9/4 (0.17\%)  & 11/5 (2.08\%) & ~              \\ 
			\hline
			8  & 20.30 & 12/5 (0.10\%) & ~             & ~              \\ 
			\hline
			9  & 21.20 & 5/2 (0.35\%)  & ~             & ~              \\ 
			\hline
			10 & 23.05 & 11/4 (0.81\%) & 8/3 (2.27\%)  & ~              \\ 
			\hline
			11 & 25.35 & 3/1 (0.00\%)  & ~             & ~              \\ 
			\hline
			12 & 28.20 & 10/3 (0.12\%) & 17/5 (1.86\%) & ~              \\ 
			\hline
			13 & 31.80 & 15/4 (0.35\%) & 19/5 (0.97\%) & ~              \\ 
			\hline
			14 & 33.98 & 4/1 (0.53\%)  & ~             & ~              \\ 
			\hline
			15 & 36.25 & 17/4 (0.94\%) & 13/3 (1.01\%) & 21/5 (2.12\%)  \\ 
			\hline
		\end{tabular}
	\end{table}